\documentclass[fleqn,usenatbib]{mnras}

\usepackage{newtxtext,newtxmath}

\usepackage[T1]{fontenc}

\DeclareRobustCommand{\VAN}[3]{#2}
\let\VANthebibliography\thebibliography
\def\thebibliography{\DeclareRobustCommand{\VAN}[3]{##3}\VANthebibliography}

\usepackage{graphicx}	
\usepackage{amsmath}	
\usepackage{xspace}
\usepackage{orcidlink}

\newcommand{\newacronym}[3]{%
  \newcommand{#1}{#2 (#3)\xspace%
    \renewcommand{#1}{#3\xspace}%
  }%
}

\newacronym{\GRMHD}{General Relativistic Magnetohydrodynamics}{GRMHD}
\newacronym{\MHD}{Magnetohydrodynamics}{MHD}
\newacronym{\RHS}{right hand side}{RHS}
\newacronym{\ECP}{Exascale Computing Project}{ECP}

\newcommand{\codename}[1]{\texttt{#1}}

\newcommand{\carpetx}{\codename{CarpetX}\xspace}

\newcommand{\theCode}{\codename{GRaM-X}\xspace}

\newcommand{\EinsteinToolkit}{\codename{Einstein Toolkit}\xspace}

\title[3D GRMHD CCSN simulations on GPUs]{\centering 3D full-GR simulations of magnetorotational core-collapse supernovae on GPUs: A systematic study of rotation rates and magnetic fields}
\author[S. Shankar et al.]{
Swapnil Shankar,$^{1}$\thanks{E-mail: swapnilshankar1729@gmail.com, sshanka5@utk.edu}
Philipp M\"{o}sta,$^{2}$
Roland Haas\,\orcidlink{0000-0003-1424-6178}\,$^{3,4,5}$ and
Erik Schnetter\,\orcidlink{0000-0002-4518-9017}\,$^{6,7,8}$
\\
$^{1}$Department of Physics and Astronomy, University of Tennessee Knoxville, Knoxville, TN 37996, USA\\
$^{2}$GRAPPA, 
Anton Pannekoek Institute for Astronomy and Institute of
High-Energy Physics, University of Amsterdam, Science Park 904, \\ 1098 XH Amsterdam, The Netherlands\\
$^{3}$Department of Physics \& Astronomy, University of British Columbia, Vancouver, Canada \\
$^{4}$National Center for Supercomputing applications, University of Illinois, Urbana, Illinois, USA \\
$^{5}$Department of Physics, University of Illinois, Urbana, Illinois, USA \\
$^{6}$Perimeter Institute for Theoretical Physics, Waterloo, Ontario, Canada \\
$^{7}$Department of Physics and Astronomy, University of Waterloo, Waterloo, Ontario, Canada \\
$^{8}$Center for Computation \& Technology, Louisiana State University, Baton Rouge, Louisiana, USA
}

\date{Accepted XXX. Received YYY; in original form ZZZ}

\pubyear{\the\year{}}

\begin{document}
\label{firstpage}
\pagerange{\pageref{firstpage}--\pageref{lastpage}}
\maketitle

\begin{abstract}
We present a series of fully three-dimensional, dynamical-spacetime general relativistic magnetohydrodynamics (GRMHD) simulations of core-collapse supernovae (CCSNe) for a progenitor of zero-age-main-sequence (ZAMS) mass $25\, M_\odot$. We simulate a total of 12 models for simulation times in the range $190-260\, \mathrm{ms}$ to systematically study the effect of rotation rates and magnetic fields on jet formation via the magnetorotational mechanism. We have performed simulations on OLCF's Frontier using the new GPU-accelerated dynamical-spacetime GRMHD code \theCode for magnetic fields $B_0 = (10^{11}, 10^{12})\, \mathrm{G}$ and rotation rates $\Omega_0 = (0.14, 0.5, 1.0, 1.5, 2.0, 2.5)\, \mathrm{rad/s}$. We always resolve the entire region containing the shock with a resolution of at least $1.48\, \mathrm{km}$. We find that models with $B_0=10^{11}\, \mathrm{G}$ fail to explode, while those with $B_0=10^{12}\, \mathrm{G}$ show a wide range of jet morphologies and explosive outcomes depending on the rotation rate. Models with $B_0=10^{12}\, \mathrm{G}$ and $\Omega_0=(1.0,1.5)\, \mathrm{rad/s}$ form jets that bend sideways, giving the ejecta a more spherical character, and possibly representing explosions that \textit{appear} neutrino-driven even though they are magnetorotationally-driven. Models with $B_0=10^{12}\, \mathrm{G}$ and $\Omega_0\geq2.0\, \mathrm{rad/s}$ show ejecta velocities $\gtrsim15000\, \mathrm{km/s}$, making them suitable candidates for broad-lined type Ic supernova progenitors. This work represents the largest set of 3D general-relativistic GRMHD simulations studying magnetorotational supernovae in full GR and demonstrates the potential of systematic studies with GPU-accelerated 3D simulations of CCSNe.

\end{abstract}




\section{Introduction}
Massive stars ($\gtrapprox 10\, M_{\odot}$) end their lives in the collapse of their iron cores. This gravitational collapse results in the formation of a protoneutron star (PNS) or black hole (BH) at the center while releasing  $\mathcal{O}(10^{53})\, \mathrm{erg}$ of energy. Most of this energy ($\sim99\%$) escapes in the form of neutrinos, but a fraction of these neutrinos can couple to the star and power a core-collapse supernova (CCSN) explosion. In this "neutrino-driven" explosion mechanism, neutrinos emitted from the cooling PNS deposit energy behind the stalled shock ("gain region") and revive it for a successful explosion. Typical neutrino-driven CCSNe have explosion energies of $\mathcal{O}(10^{51})\, \mathrm{erg}$ \citep{Burrows2021}. 

While this likely drives the vast majority of CCSNe, a small fraction of observed supernovae have kinetic energies $\sim10$ times that of neutrino-driven CCSNe ($\sim10^{52}\, \mathrm{erg}$) and broad spectral lines indicating relativistic outflows ($15000-30000\, \mathrm{km/s}$). These supernovae are classified as broad-lined type-Ic supernovae (SNe Ic-bl) or \textit{hypernovae} (short for hyper-energetic supernovae) and sometimes are accompanied by a long Gamma Ray Burst (GRB) \citep{Modjaz2011, hjorth_bloom_2012, Cano2017}. The kinetic energies of the ejecta and broadened spectral lines observed in these SNe cannot be explained by the neutrino mechanism alone \citep[e.g.][]{Burrows2021}. The leading mechanism to explain these explosions is jet-driven or magnetorotational mechanism. In this magnetorotational mechanism, rapid rotation of the collapsed core ($\mathcal{O}(1)\, \mathrm{ms}$) and magnetar-strength magnetic fields ($\sim10^{15}\, \mathrm{G}$) can launch a jet-like outflow along the rotation axis, which is able to explain the high kinetic energies and velocities observed in SNe Ic-bl or hypernovae.

Previous 2D studies have indicated that rapid rotation in the pre-collapse star with periods $\lessapprox 4\, \mathrm{s}$~\citep{Ott_2006, Burrows_2007} is an important requirement for the magnetorotational mechanism to operate. The magnetic fields at the end stage of the evolution of massive stars just before collapse are likely weak ($<<10^{15}\, \mathrm{G}$). In order for the magnetic field strength to reach magnetar levels after collapse, an efficient mechanism is needed which can amplify the "seed" magnetic fields. The magnetorotational instability \citep[MRI, ][]{Balbus1991, Akiyama2003} has been proposed as a possible mechanism for such amplification. The MRI operates on differential rotation in the star and has been shown to grow field exponentially fast \citep{Balbus1991}. However, resolving the wavelength of the fastest-growing mode of MRI requires spatial resolution of $\lessapprox50$-$100\, \mathrm{m}$ within the post-bounce core. This is about $\sim 5x$ higher than what can be resolved in typical production-level CCSN simulations ($\sim300\, \mathrm{m}$). \cite{Moesta2015} resolved the MRI in dedicated global simulations and showed that this mechanism can amplify realistic initial magnetic fields of $\sim10^{10}\, \mathrm{G}$ to magnetar strength fields ($\sim10^{15}\, \mathrm{G}$) and lead to the global field structure needed to launch jet-like outflows. A common approach in the community has been to not resolve the MRI and instead start with likely unphysically high seed magnetic fields of $10^{12}-10^{13}\, \mathrm{G}$ on the pre-collapse progenitors to mimic the field resulting from the MRI.

The rotation profile is often also initially imposed onto the progenitor core using a parametric model, as final rotation profiles from 1D stellar evolution simulations performed with codes such as MESA~\citep{Paxton2011, Paxton2013} are likely not reliable as inputs for 3D CCSN simulations. In practice, many groups often choose a rotation profile favorable for the magnetorotational mechanism and assume that such rotation profiles can be realistically reached in at least a small fraction of progenitors, which is not entirely unreasonable given that at most 1\% of all CCSNe are hypernovae~\citep{Woosley_2006, Winteler_2012}.

Given these uncertainties in the initial magnetic field and rotation profiles of massive stars prior to collapse, it is important to study the effects of different rotation profiles and magnetic fields on the explosion dynamics. 3D studies have so far been limited to only a few models due to the computational cost of these simulations \citep{Moesta_2018, Kuroda_2020, Obergaulinger_2021, Bugli_2021, Powell_2023, Shibagaki_2024}. 2D studies have allowed to probe the parameter space more easily but are limited as MHD instabilities that can fundamentally change the dynamics of jet formation and propagation can only be captured in full 3D~\citep{Moesta2014}. The state of the art currently is defined by a few groups that have done at most 4 simulations in 3D: \cite{Shibagaki_2024} have performed 3D GRMHD simulations of 4 models parametrically varying the rotation rates and magnetic fields for a $20\, M_\odot$ progenitor. \cite{Obergaulinger_2021} have studied 4 different pre-collapse magnetic field configurations for a $35\, M_\odot$ progenitor using special-relativistic MHD simulations. \cite{Kuroda_2020} have computed 3 models varying rotation rate and magnetic field for a $20\, M_\odot$ progenitor with 3D GRMHD simulations. \cite{Powell_2023}, \cite{Bugli_2021}, \cite{Moesta_2018}, and \cite{Halevi_2018} have studied the effect of different magnetic field configurations. 

In this work, we present 12 magnetorotational core-collapse supernova simulations using a progenitor of zero-age-main-sequence mass $25\,M_{\odot}$. Our aim is to systematically study the conditions under which jets can be successfully launched and to quantitatively study the properties of the explosion as a function of initial rotation rates and magnetic fields. We choose the same progenitor and systematically vary the initial rotation rates and magnetic fields and always resolve the shock with a resolution of at least $1.48\, \mathrm{km}$ for all models. To the best of our knowledge, this work represents the largest set of 3D GRMHD simulations and the most comprehensive study of rotation rates and magnetic fields for a single progenitor. This has been enabled by our newly-developed GPU-accelerated dynamical-spacetime ideal-GRMHD code \theCode~\citep{Shankar2023}. \theCode has been developed for the \EinsteinToolkit framework~\citep{Loffler_2012, roland_haas_2024_zenodo} and relies on the new adaptive mesh-refinement (AMR) drive \carpetx~\citep{schnetter_2022_zenodo, Shankar2023, Kalinani_2025}. We have performed the simulations at Oak Ridge Leadership Computing Facility's Frontier. The total cost of all simulations presented is $\sim35000$~node-hours demonstrating that 3D GRMHD parameter studies are feasible now for magnetorotationally-driven CCSNe. 

The outline of the paper is as follows. In section~\ref{sec:methods}, we discuss the initial conditions used for the simulations (progenitor, rotation profile, magnetic fields) and the numerical setup and methods. In section~\ref{sec:results}, we present the results of the simulations including postbounce dynamics, shock propagation, accretion onto the PNS and properties of the explosion. We present the conclusions in section~\ref{sec:conclusions}.

\section{Methods}
\label{sec:methods}

\subsection{Pre-collapse initial conditions}

\subsubsection*{Progenitor}
We take a $25\, M_\odot$ zero-age main sequence (ZAMS) mass pre-collapse progenitor model from \cite{Aguilera-Dena_2020}. The model provides 1D profiles of density, temperature, electron fraction and velocities which we map to the 3D grid of \theCode. We then add a range of parametric rotation profiles and magnetic field configurations on top of this progenitor model to perform 3D magnetorotational CCSN simulations. \\

\cite{Halevi_2023} have simulated this progenitor model in 1D through core collapse and until $767\, \mathrm{ms}$ post bounce using the code GR1D~\citep{OConnor_2010}. This progenitor has a core-compactness parameter, $\xi_{2.5} = 0.47$, greater than 0.45 and thus this progenitor is expected to be difficult to successfully explode based on the value of this traditional 'explodability' indicator~\citep{OConnor_2011, Halevi_2023}. In the absence of magnetic fields, the 1D simulations performed in \cite{Halevi_2023} fail to reach an explosion, with the shock reaching a maximum radius of $\sim140\, \mathrm{km}$ at $\sim80\, \mathrm{ms}$, and then stalling and receding. However, black hole is not formed in the simulations either, and the authors conclude that this model may be representative of a successful neutrino-driven explosion at later times. In this paper, we simulate this model in 3D, with a range of rotation profiles and magnetic fields imposed on the progenitor.

\subsubsection*{Rotation profile and magnetic fields}
We set up an initial differential rotation profile using the rotation law of \cite{Takiwaki_2011} where the angular velocity $\Omega$ is given by
\begin{equation}
    \Omega(r, z) = \Omega_0 \frac{r_0^2}{r^2 + r_0^2} \frac{z_0^4}{z^4 + z_0^4},
\end{equation}
and where $r$ is the distance from the rotation axis in the equatorial plane and $z$ the distance from the equatorial plane. We choose parameters $r_0 = 500\, \mathrm{km}$  and $z_0 = 1000\, \mathrm{km}$. The parameter $r_0$ controls the degree of differential rotation and our choice of $r_0 = 500\, \mathrm{km}$ implies moderate differential rotation. In order to study the effect of rotation rate on the magnetorotational mechanism, we vary $\Omega_0$ from $0.014\, \mathrm{rad/s}$ to $2.5\, \mathrm{rad/s}$ which correspond to pre-collapse periods of $44.8\, \mathrm{s}$ to $2.1\, \mathrm{s}$, respectively. The lower limit of $\Omega_0 = 0.014\, \mathrm{rad/s}$ is determined from the 1D stellar evolution model, based on the peak angular velocity extracted from the progenitor’s 1D profile.

We set up the initial pre-collapse magnetic field using a vector potential of the form
\begin{equation}
    A_r = 0, \\
    A_{\theta} = 0,\\
    A_{\phi} = B_0  \frac{r_0^3}{r^3+r_0^3} r \mathrm{sin} \theta
\end{equation}
where $A_r$, $A_{\theta}$ and $A_{\phi}$ are the $r$, $\theta$ and $\phi$ components of the magnetic vector potential. We choose $r_0 = 2000\, \mathrm{km}$ which corresponds to the approximate radius of the pre-collapse core. We perform simulations for values of $B_0 = 10^{11}\, \mathrm{G}$ and $B_0 = 10^{12}\, \mathrm{G}$.

\subsection{Magnetorotational CCSN simulations}

We simulate a total of 12 models by varying the rotation rate parameter $\Omega_0$ and the initial magnetic field strength $B_0$ for the same $25\, M_\odot$ ZAMS progenitor. The models are described in Table~\ref{tab:model_list}. We use notation \texttt{B11} (\texttt{B12}) to collective refer to all models with $B_0=10^{11}\, \mathrm{G}$ ($B_0=10^{12}\, \mathrm{G}$).

\begin{table}
    \centering
    \begin{tabular}{|c|c|c|c|c|c|c|}
        \hline
        Model & $\Omega_0$~($\mathrm{rad/s}$) & $B_0$~(G) &  & Model & $\Omega_0$~($\mathrm{rad/s}$) & $B_0$~(G) \\ \hline
        \texttt{R01B11}    & 0.14    & $10^{11}$ &  & \texttt{R01B12}    & 0.14    & $10^{12}$     \\ \hline
        \texttt{R05B11}    & 0.5     & $10^{11}$ &  & \texttt{R05B12}    & 0.5     & $10^{12}$     \\ \hline
        \texttt{R10B11}    & 1.0     & $10^{11}$ &  & \texttt{R10B12}    & 1.0     & $10^{12}$     \\ \hline
        \texttt{R15B11}    & 1.5     & $10^{11}$ &  & \texttt{R15B12}    & 1.5     & $10^{12}$    \\ \hline
        \texttt{R20B11}    & 2.0     & $10^{11}$ &  & \texttt{R20B12}    & 2.0     & $10^{12}$     \\ \hline
        \texttt{R25B11}    & 2.5     & $10^{11}$ &  & \texttt{R25B12}    & 2.5     & $10^{12}$     \\ \hline
    \end{tabular}
    \caption{Summary of models simulated.}
    \label{tab:model_list}
\end{table}

\subsubsection*{Spacetime and GRMHD evolution}
We perform the simulations using the new GPU-accelerated dynamical-spacetime GRMHD code \texttt{GRaM-X} as part of the \texttt{Einstein Toolkit} framework. We use the Z4c formulation~\citep{Milton:2011, hilditch2013} for solving Einstein's equations and the Valencia formulation \citep{Marti:1991wi,Banyuls:1997zz,Ibanez:2001:godunov,Font:2007zz} within a finite-volume framework to follow the ideal-GRMHD evolution. \theCode provides support for finite-temperature tabulated equations of states (EoS) and in simulations presented here, we use the $K_0 = 220\, \mathrm{MeV}$ variant of the EoS of ~\cite{Lattimer:1991nc} which is commonly referred to as \textit{LS220}. We use the 5th order WENO5 (Weighted Essentially Non-Oscillatory) reconstruction method~\citep{shu:98} with fallback to 2nd order TVD (Total Variation Diminishing) reconstruction~\citep{toro:99} in the regions where WENO5 gives unphysical values (such as in the regions of extreme dynamics) and the HLLE Riemann solver~\citep[Harten-Lax-van Leer-Einfeldt,][]{Einfeldt:1988og, Harten:1983hr, Gammie:2003rj} to solve the Riemann problem at cell interfaces. We use an approximate neutrino treatment using the M0 scheme as described in \cite{Radice_2016, Radice_2018}. While the M0 transport approximation is not sufficient to study the detailed composition of the ejecta and nucleosynthesis, our focus in this paper is on exploring a large number of models to pave the way for 3D parameter studies of magnetorotational CCSNe. As  neutrinos are energetically subdominant for the explosion dynamics of magnetorotational CCSNe, we choose this computationally less costly approach here. We are currently performing detailed follow-up simulations using the M1 scheme presented in \cite{Radice:2022} that we used in the context of neutron-star mergers~\citep{curtis_2023}.

\subsubsection*{Grid Setup}

We start with the same numerical setup for each model. We initialize the simulations with 4 AMR levels with basegrid resolution of $94\, \mathrm{km}$ and boundaries at $\pm 6000\, \mathrm{km}$ in each direction. Refined levels have extents $\pm 1890\, \mathrm{km}$, $\pm 760\, \mathrm{km}$ and $\pm 380\, \mathrm{km}$ with a factor 2 resolution difference between each refinement level. Our base grid has $128^3$ cells, while the refined levels have $80^3$, $64^3$ and $64^3$ cells, respectively. During core collapse, we add 5 more levels progressively when the central density exceeds $8\times10^{10}$, $3.2\times10^{11}$, $1.28\times10^{12}$, $5.12\times10^{12}$, and $2.048\times10^{13}\, \mathrm{g/cm}^3$, respectively. These additional levels have extents $\pm 190\, \mathrm{km}$, $\pm 120\, \mathrm{km}$, $\pm 60\, \mathrm{km}$, $\pm 30\, \mathrm{km}$ and $\pm 20\, \mathrm{km}$ and with the factor 2 resolution difference between levels, our finest level has a resolution of $370\, \mathrm{m}$. 

During postbounce evolution we track the expanding shock wave in successful explosions by adjusting the boundaries of refinement levels. We always contain the shock in AMR level 6 with resolution $1.48\, \mathrm{km}$. The initial extent of level 6 is $\pm$60 km, but we dynamically increase this extent to make sure the shock is contained within it. In turn, the simulations become progressively more computationally expensive as the shock expands. For example, the shock expands to a radius of $720\, \mathrm{km}$, $942\, \mathrm{km}$ and $1500\, \mathrm{km}$ in $x$, $y$ and $z$ at the end of the simulation for model \texttt{R25B12}. As a result, AMR level 6 contains a total of $1080\times1380\times2140 \simeq 1472^3$~cells (starting from the initial $80^3$~cells), with other levels also expanding. This is a key difference to other studies~\citep{Bugli_2021, Obergaulinger_2021, Shibagaki_2024}, where the expanding shockwave is less well resolved as it propagates to larger radii.   


\subsubsection*{Conservative-to-primitive transformation}
We use a 3D Newton-Raphson \citep[\textit{3DNR}, ][]{cerda_duran:2008} as the primary conservative-to-primitive transformation method (referred to as "con2prim" hereafter). If this primary method fails, we use the method of Newman \& Hamlin \citep[\textit{Newman}, ][]{Newman_Hamlin:2014} as a fallback. However, it is possible that in the extreme regions of the simulations, both \textit{3DNR} and \textit{Newman} fail for individual points. We have found that this usually happens in cells with high magnetic fields and low densities because unphysical solutions can be obtained in these regions of extreme magnetization due to limitations of the ideal GRMHD approximation. In case of such failures, we average the temperature from neighboring cells and self-consistently perform another con2prim with \textit{Newman} using the calculated average temperature as the given temperature (i.e we only iterate over $\rho$, as opposed to the standard method of iterating over $\rho$ and temperature). We find that these failures typically only happen at a few points in the regions of extreme dynamics and each time resolve after a few iterations. 

\subsubsection*{Shock tracking}
Since we expand the refinement levels to ensure that the entire region inside the shock is covered with a resolution of at least $1.48\, \mathrm{km}$, we need to use a mechanism to accurately determine the three-dimensional surface of the shock. We employ two different methods for shock tracking, one more reliable at earlier times and the other more reliable at later times. At earlier times ($t_{pb} \lesssim100\, \mathrm{ms}$), we track the shock by locating the boundary where the mass fraction of heavy elements, $X_h$, is less than 0.5~\citep{Ott_2013}. This method works because $X_h$ is typically $\lesssim0.1$ inside the shock and $\gtrsim0.9$ outside the shock.  However, at later times, this method is not able to capture the shape of the entire shock front accurately. In this case instead, we use the gradient of the pressure to track the shock. We calculate the pressure gradient, $\nabla P$, for all points in the simulation domain and find that the condition $\nabla P > 0.1$ is able to accurately determine the 3D shock surface. 

\subsubsection*{Diffusivity for the magnetic field}
In the postbounce phase, we find that if the shock remains stalled for a few ms, undesirable high-frequency oscillations can develop in the magnetic field just outside the shock. These oscillations cause magnetohydrodynamic instabilities and were also seen in~\cite{Moesta2014}. To avoid this issue, we add diffusivity to the magnetic field \textit{outside} the shock. We achieve this using a modified Ohm's Law for the electric field calculation in our constrained transport implementation, given by~\citep{Moesta2015} 
\begin{equation}
    \mathbf{E} = -\mathbf{v} \times \mathbf{B} + \eta \mathbf{J}, \\
    \mathbf{J} = \nabla \times \mathbf{B}
\end{equation}
where $\mathbf{E}$ is the electric field, $\mathbf{B}$ is the magnetic field, $\mathbf{v}$ is the 3-velocity and $\mathbf{J}$ is the 3-current density. We set $\eta = 0$ inside the shock and $\eta = 1$ outside the shock. To enable a smooth transition between the values of $\eta$ inside and outside the shock, we use the convolution operation in the form of nearest neighbour averaging. We first set the values of $\eta$ strictly 0 or 1 and then average $\eta$ with neighbors along $3\times 3 \times 3$ cubes. This leads to a smooth transition of $\eta$ from 0 (inside shock) to 1 (outside shock) on a length scale of $\sim 2$ km at the shock interface. It should be noted that we do not use a simple spherical model for setting $\eta$ as the shock can be highly aspherical.

\subsubsection*{Treatment of regions of high magnetization}
We define magnetization as $\sigma = b^2/\rho$, where $\rho$ is the fluid rest-mass density and $b^2$ is the magnetic field in the fluid rest frame. We find that in the regions of extreme dynamics the value of $\sigma$ can sometimes become very high leading to numerical problems due to violations of the ideal MHD approximation. Following similar approaches in simulations of black-hole accretion disks and their jets, we limit $\sigma$. If $\sigma$ in any cell rises above the value $\sigma_{\mathrm{limit}}$, we recalculate the density in the cell using $\rho = b^2/\sigma_{\mathrm{limit}}$. We then update the values of pressure and internal energy based on the new $\rho$. Therefore, limiting $\sigma$ is equivalent to artificial mass injection. We generally use $\sigma_{\mathrm{limit}} = 15$ for all simulations. However, the regions just behind the shock in models \texttt{R20B12} and \texttt{R25B12} reach extremely high magnetizations at late times ($\gtrsim180\, \mathrm{ms}$ postbounce) in which case we use $\sigma_{\mathrm{limit}} = \mathcal{O}(1)$ for stable evolution \citep[for more details, see appendix C of ][]{Ressler_2017}. We face this issue in these most extreme models because of the high resolution we have within the entire shocked region as we find that using lower resolution increases numerical dissipation and thus leads to less extreme magnetization.


\section{Results}
\label{sec:results}

We start our simulations from the precollapse progenitor star with the initial conditions described in Section \ref{sec:methods} and Table \ref{tab:model_list}. During collapse, there is a steep rise in central density from $\sim10^9\, \mathrm{g/cm}^3$ to nuclear density ($\sim10^{14}\, \mathrm{g/cm}^3$), accompanied by a steep drop in the central lapse $\alpha$ from $\sim1$ to $\sim0.84$.  We define core bounce as the time at which the central density reaches $8\times10^{13}\, \mathrm{g/cm}^3$. Core bounce is reached at time $t_b \sim 164-167\, \mathrm{ms}$ for the different models, and it depends slightly on the initial magnetic field and rotation parameter choice.

\subsection{Postbounce dynamics}

\begin{figure*}
 	\includegraphics[width=1.0\textwidth]{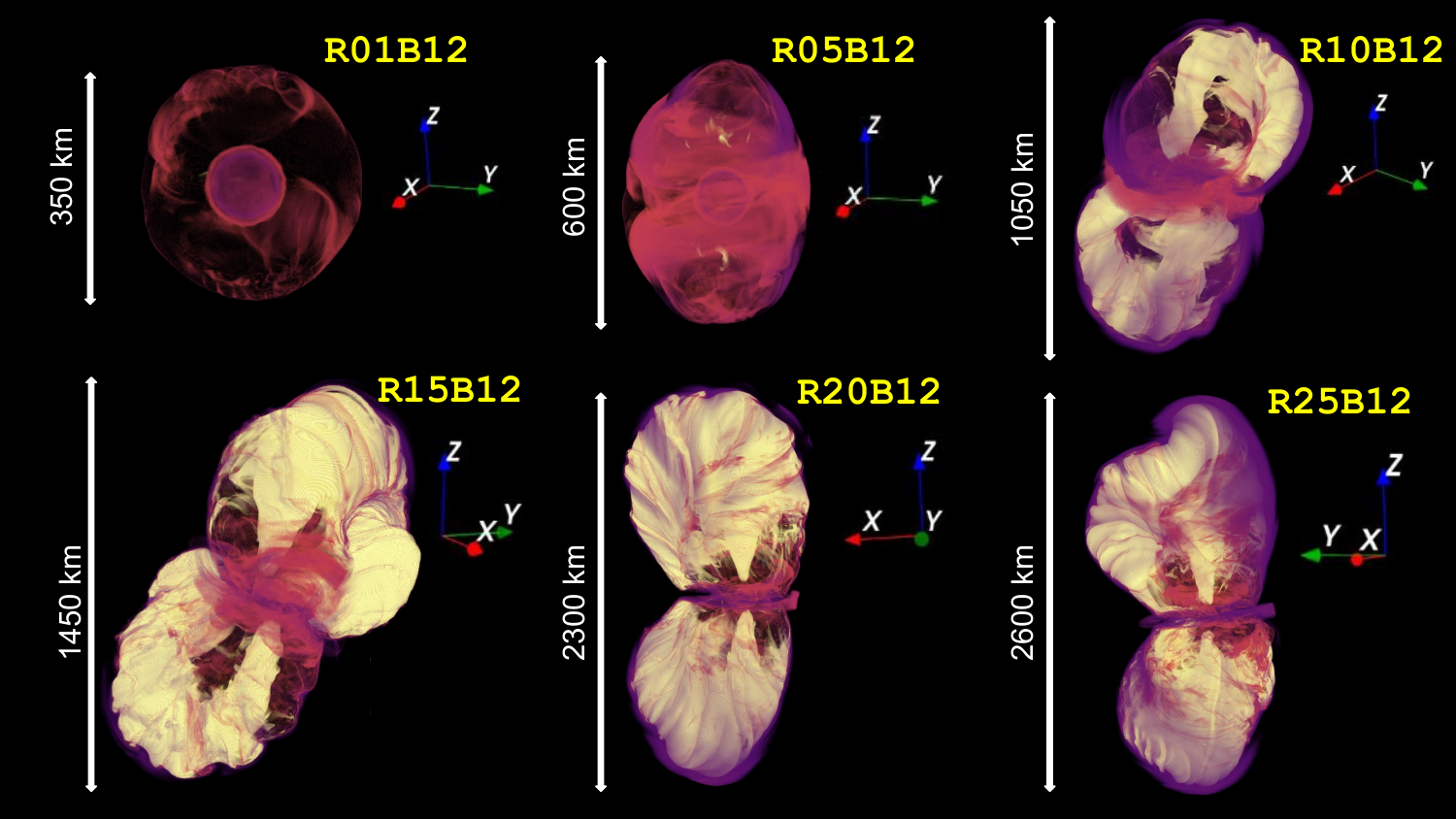}
    \caption{Volume rendering of entropy for all \texttt{B12} models at $200\, \mathrm{ms}$ postbounce, except for model \texttt{R25B12} which is shown at $190\, \mathrm{ms}$ postbounce. The scale on the left indicates the polar extent of the shock surface for each model. Violet regions correspond to entropy of $\sim7k_b\, \mathrm{baryon}^{-1}$, reddish-orange corresponds to entropy of $\sim10k_b\, \mathrm{baryon}^{-1}$ and bright-yellow corresponds to material with high entropy of $\gtrsim16k_b\, \mathrm{baryon}^{-1}$.} 
    \label{fig:volume_rendering}
\end{figure*}

\begin{figure*}
 	\includegraphics[width=0.75\textwidth]{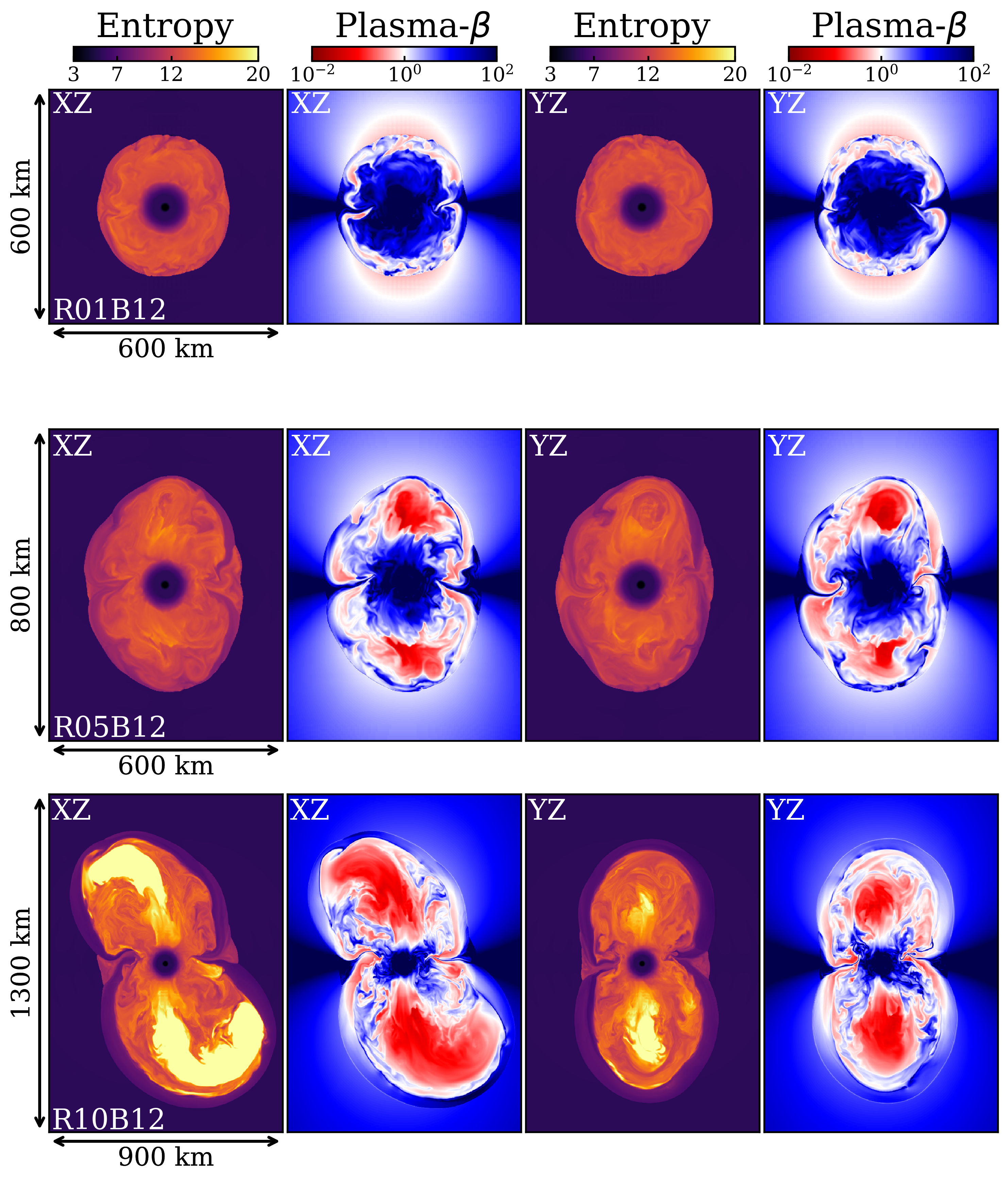}
    \caption{2D slices of entropy (in $k_b\, \mathrm{baryon}^{-1}$) and plasma-$\beta$ for models \texttt{R01B12}, \texttt{R05B12} and \texttt{R10B12} at $200\, \mathrm{ms}$ postbounce. The two columns on the left show entropy and plasma-$\beta$ in the $xz$-plane for $y=0$ and the two columns on the right show these quantities in the $yz$-plane for $x=0$. In the entropy panels, dark-purple regions correspond to low-entropy material ($\sim4k_b\, \mathrm{baryon}^{-1}$), violet to the shock surface ($\sim5k_b\, \mathrm{baryon}^{-1}$), reddish-orange to high-entropy material ($\sim10-15 k_b\, \mathrm{baryon}^{-1}$) and bright-yellow to very high entropy material ($\gtrsim 20k_b\, \mathrm{baryon}^{-1}$).  In the plasma-$\beta$ panels, red regions indicate highly-magnetized material (dominated by magnetic pressure, $\beta<1$). We note that we change the coordinate axis ranges for the different models to match the progressing shock expansion. The entire domain shown here as well as in figures~\ref{fig:andes_set2}, \ref{fig:andes_set3} and \ref{fig:andes_set4} has a resolution of at least $1.48\, \mathrm{km}$.} 
    \label{fig:andes_set1}
\end{figure*}

\begin{figure*}
 	\includegraphics[width=0.75\textwidth]{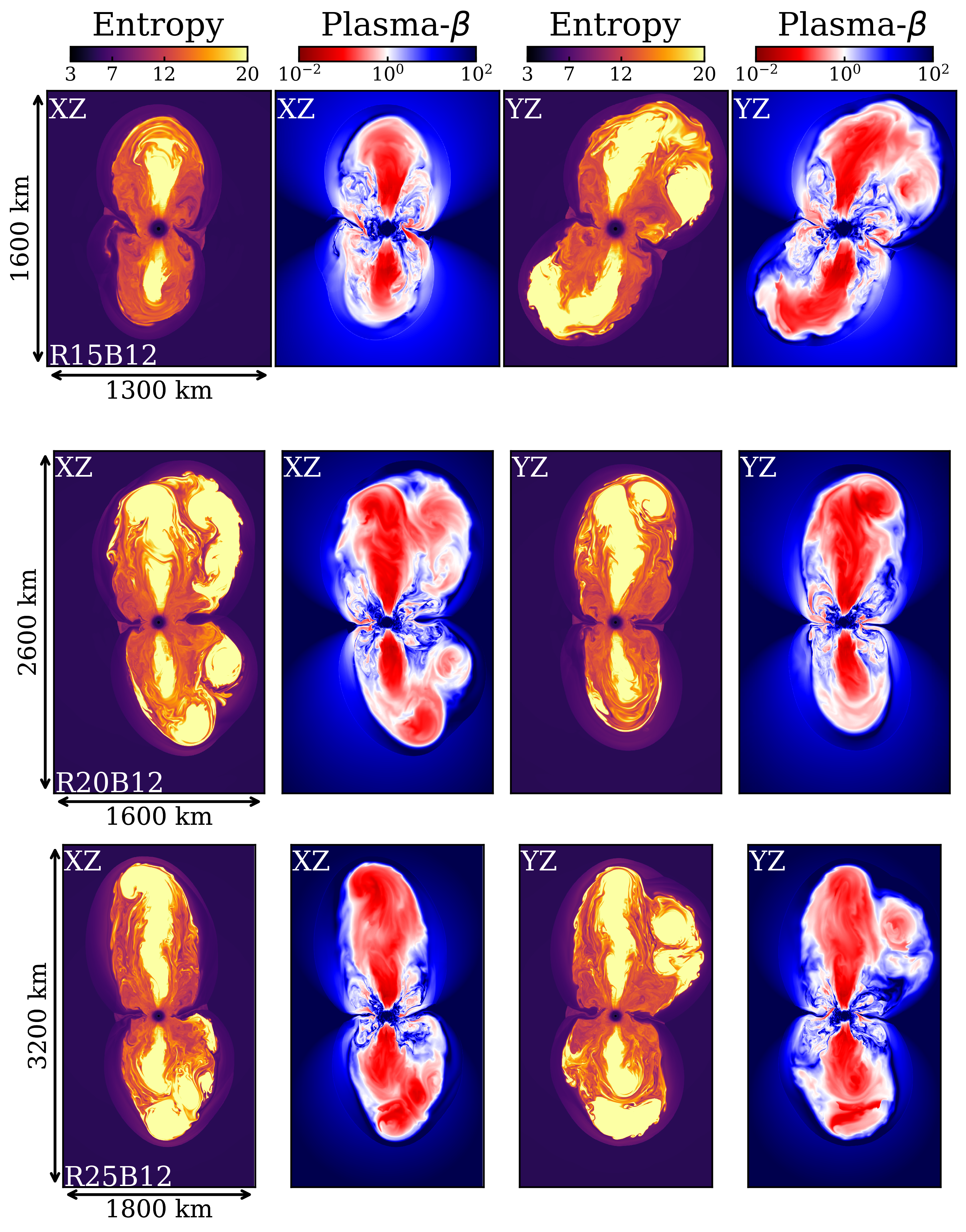}
    \caption{ As in Fig.~\ref{fig:andes_set1}, but for models \texttt{R15B12}, \texttt{R20B12} and \texttt{R25B12}. Models \texttt{R15B12} and \texttt{R20B12} are shown at $200\, \mathrm{ms}$ postbounce, while model \texttt{R25B12} is shown at $190\, \mathrm{ms}$ postbounce.} 
    \label{fig:andes_set2}
\end{figure*}

\begin{figure*}
 	\includegraphics[width=0.75\textwidth]{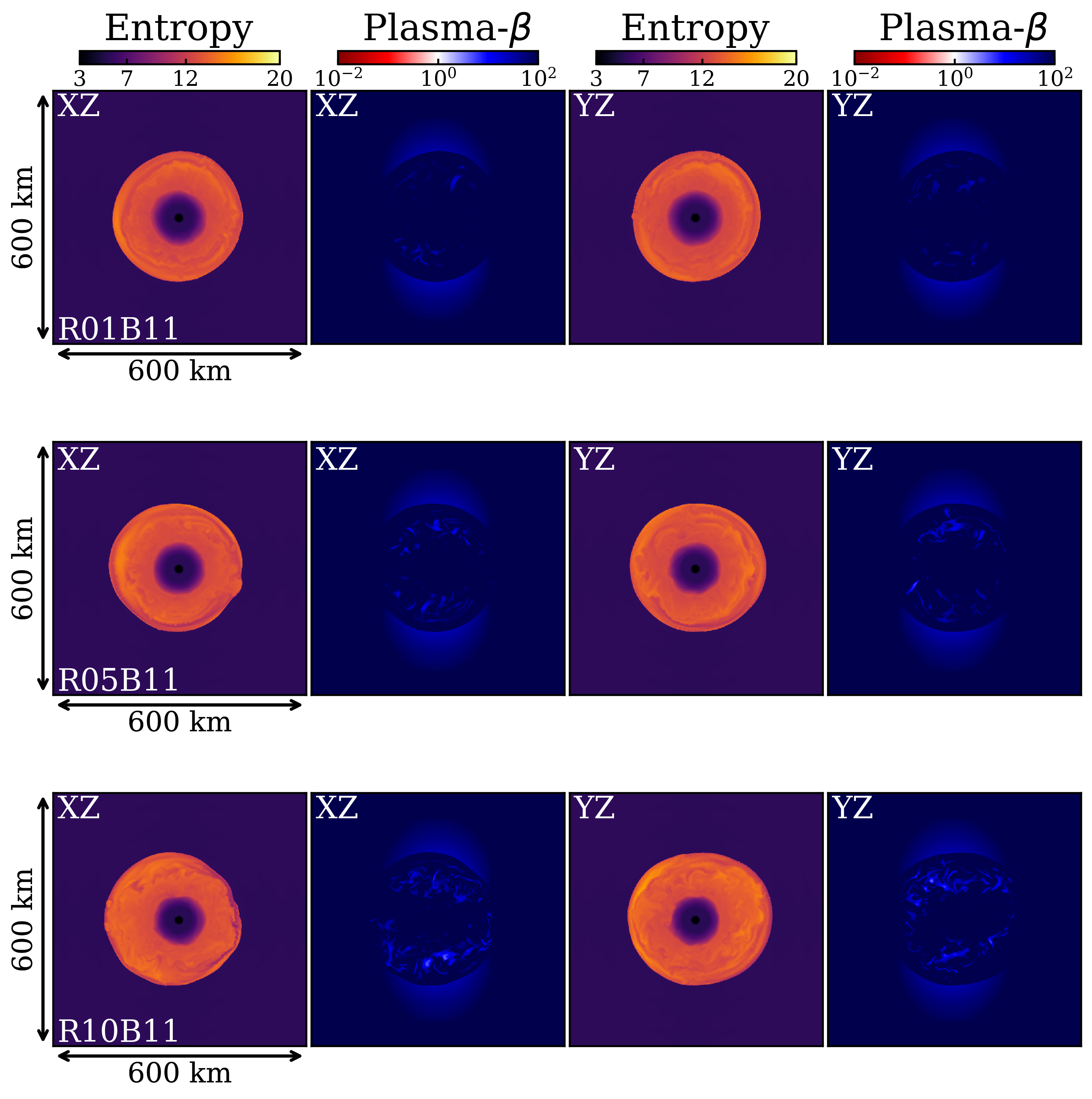}
    \caption{ As in Fig.~\ref{fig:andes_set1}, but for models \texttt{R01B11}, \texttt{R05B11} and \texttt{R10B11} at $200\, \mathrm{ms}$ postbounce. } 
    \label{fig:andes_set3}
\end{figure*}

\begin{figure*}
 	\includegraphics[width=0.75\textwidth]{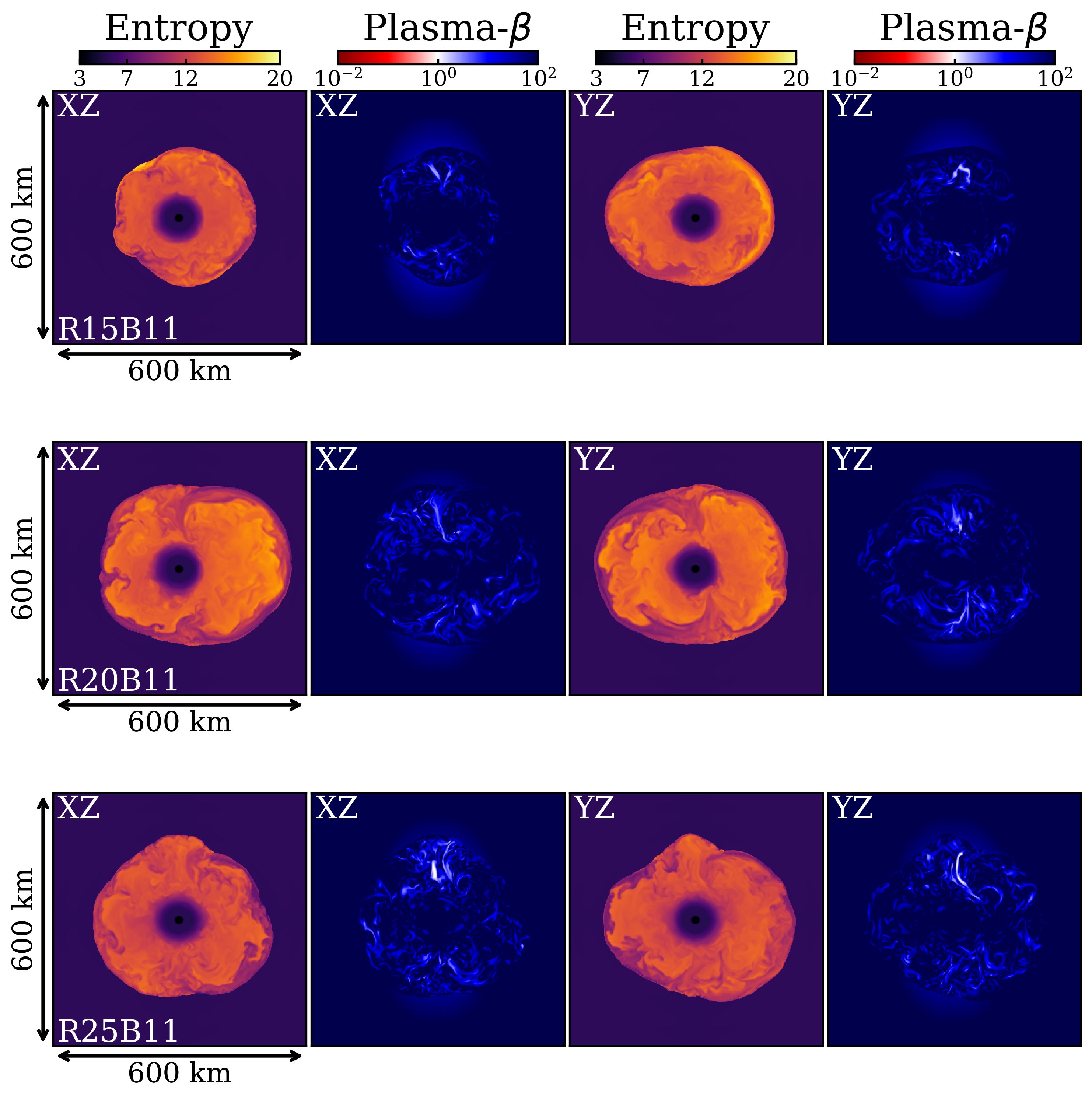}
    \caption{ As in Fig.~\ref{fig:andes_set1}, but for models \texttt{R15B11}, \texttt{R20B11} and \texttt{R25B11} at $200\, \mathrm{ms}$ postbounce. } 
    \label{fig:andes_set4}
\end{figure*}

All models show very similar initial shock evolution until $\sim80\, \mathrm{ms}$ postbounce. A spherical magnetohydrodynamic shock is launched at bounce but stalls at $\sim10\, \mathrm{ms}$ and at a radius of $\sim110\, \mathrm{km}$. This stalling is caused as the shock wave loses energy running into the accreting material of the core and surrounding star and loses energy dissociating iron group elements. Nearly $15\, \mathrm{ms}$ after bounce, the toroidal component of the magnetic field, $B_{\mathrm{tor}}$, has grown to $\sim10^{14}-10^{15}\, \mathrm{G}$ for the \texttt{B11} models and $\gtrsim10^{15}\, \mathrm{G}$ for the \texttt{B12} models. Thus, all the models reach magnetar-strength magnetic fields. The peak value of $B_{\mathrm{tor}}$ depends on the rotation rate, with models having larger rotation rate showing larger values of peak $B_{\mathrm{tor}}$, as expected from rotational winding~\citep{Shibata_2006}. Among the \texttt{B12} models, the value of $B_{\mathrm{tor}}$ at $15\, \mathrm{ms}$ postbounce ranges from $1\times10^{15}\, \mathrm{G}$ in model \texttt{R01B12} to $4\times10^{15}\, \mathrm{G}$ in model \texttt{R25B12}, and for the \texttt{B11} models it ranges from $3\times10^{14}\, \mathrm{G}$ in model \texttt{R01B11} to $7\times10^{14}\, \mathrm{G}$ in model \texttt{R25B11}. The evolution of the specific entropy in contrast is very similar across different models until $\sim80\, \mathrm{ms}$ postbounce, with the maximum entropy ranging from $\sim6k_b\, \mathrm{baryon}^{-1}$ at bounce to $\sim12k_b\, \mathrm{baryon}^{-1}$ at $80\, \mathrm{ms}$ postbounce.

We use the plasma-$\beta$ parameter, defined as the ratio of fluid pressure and magnetic pressure, 
$\beta = P_{\mathrm{fluid}}/P_{\mathrm{magnetic}}$, to analyze how magnetic fields influence the dynamics inside the shocked region.  Regions with $\beta \lesssim 1$ are dynamically dominated by magnetic fields and presence of such regions can be a sign of jet formation. In a fully developed jet, the value of $\beta$ can reach $\lesssim10^{-2}$. We find that regions in models \texttt{R25B12}, \texttt{R20B12} and \texttt{R15B12} become magnetically-dominated early on, with regions along the rotation axis ($Z$-axis) reaching $\beta \lesssim 1$ at $\sim20-25\, \mathrm{ms}$ postbounce, while models \texttt{R10B12}, \texttt{R05B12} and \texttt{R01B12} become magnetically-dominated later, at $\sim60\, \mathrm{ms}$ postbounce. None of the \texttt{B11} models become magnetically-dominated in the first $80\, \mathrm{ms}$ of postbounce evolution. Despite these differences, all models remain nearly spherical with a radius of $\sim130-150\, \mathrm{km}$ until $\sim80\, \mathrm{ms}$ after bounce, with the shock front sloshing around in a spiral fashion indicating the presence of spiral SASI modes~\citep{Iwakami_2009, Fernandez_2010, Endeve_2012}. The shock remains largely spherical even for the strongly-rotating \texttt{B12} models as the MHD kink instability leads to a spiral deformation of the magnetically-dominated outflow from the PNS, preventing it from successfully launching a jet explosion along the rotation axis  at this early time~\citep{Moesta_2014}. 

The evolution from $80\, \mathrm{ms}$ postbounce differs significantly between models. We show volume renderings of entropy for all the \texttt{B12} models in Fig.~\ref{fig:volume_rendering} and 2D slices of entropy and plasma-$\beta$ in the $xz$- and $yz$-planes for all models in Fig.~\ref{fig:andes_set1}, \ref{fig:andes_set2}, \ref{fig:andes_set3} and \ref{fig:andes_set4}. These snapshots are shown at $200\, \mathrm{ms}$ postbounce, except for model \texttt{R25B12} which is shown at $190\, \mathrm{ms}$ postbounce. Due to the expansion of the shockwave to large radii, the simulation of the model \texttt{R25B12} becomes very expensive at late times and we chose to only simulate it only until $190\, \mathrm{ms}$ postbounce, whereas we simulate all other models up to at least $200\, \mathrm{ms}$ postbounce. We find that none of the \texttt{B11} models are able to launch a successful jet until the end of the simulated time, with the polar regions barely getting magnetically-dominated (i.e. $\beta \gtrsim 1$). The shock front sloshes around but remains almost symmetric with a radius of $\sim150-250\, \mathrm{km}$ for all the \texttt{B11} models, with the rapidly rotating ones showing slightly oblate shock fronts at later times. Thus, the \texttt{B11} models can be ruled out as candidates for prompt jet-driven explosions, but they may explode later on as a combination of delayed magnetorotational and neutrino-driven explosions, e.g. the strongly-magnetized model \texttt{S} of \cite{Obergaulinger_2021} undergoes prompt shock revival, while the weakly-magnetized model \texttt{W} undergoes shock revival after $\sim300\, \mathrm{ms}$ postbounce.

The \texttt{B12} models, on the other hand, show a wide range of outcomes depending on initial rotation rate. For the \texttt{B12} models with the slowest rotation rates (\texttt{R01B12} and \texttt{R05B12}), a jet is not launched until the end of the simulation. Model \texttt{R01B12} remains spherically symmetric until the end and $\beta$ remains $\gtrsim10^{-1}$. This $\beta$ is still an order of magnitude lower than any of the \texttt{B11} models, yet it fails to launch a successful jet. $\beta$ for model \texttt{R05B12} is as low as $2\times10^{-2}$, but even in this case a successful jet is not launched until the end of the simulation. This model, however, shows some degree of asymmetry in the shock evolution unlike \texttt{R01B12} which is symmetric. 

For the \texttt{B12} models with $\Omega_0 \ge 1.0\, \mathrm{rad/s}$, the shock is able to break out of the polar region of the core and develops an \textit{asymmetric} bipolar outflow along the rotation axis ($z$-axis). This outflow consists of material with high magnetic pressure ($\beta \sim 10^{-2}$) and high entropy ($\gtrapprox 15-20 k_b\, \mathrm{baryon}^{-1}$). The dynamics in the later phases of the simulation ($t_b > 150\, \mathrm{ms}$) for these models involves an asymmetric bipolar outflow with pockets of low-density, highly-magnetized material pushing the shock front outwards along the rotation axis. There is a continuous outflow of material from the PNS, which however is not able to propagate cleanly along the rotation axis due to deformations~\citep[arising from the kink instability, see][]{{Moesta2014}}. Instead, the outflow interacts with the existing pockets of low-density, highly-magnetized material, bends, sometimes falls back, and sometimes forms vortex-like flow patterns. This indicates that part of the linear momentum of the outflow is converted into angular momentum, which leads to a relatively slowly moving bipolar outflow instead of a clean jet. The distortion of the jet and bipolar shock surface is much more pronounced for lower rotation rates and can be clearly seen in models \texttt{R10B12} and \texttt{R15B12} in Fig.~\ref{fig:andes_set1} and Fig.~\ref{fig:andes_set2}, where the entire lobes in the polar directions get tilted. This is plausible as the jets produced in lower rotation rate models are weaker and tilt away from the rotation axis more due to not be able to push the shockwave outwards in the polar direction. Portions of the jetted outflows also fall back for the most strongly rotating models \texttt{R20B12} and \texttt{R25B12}, but in this case the outflow is strong enough that the shock front does not fully tilt and keeps steadily moving outwards along rotation axis.

In summary, we find that all \texttt{B11} models and the slowest rotating \texttt{B12} model (i.e. \texttt{R01B12}) do not lead to jet formation and are not able to successfully explode within the simulated time. Model \texttt{R05B12} has not formed a successful jet until the end of the simulation, but shows strong signs of jet formation ($\beta\sim10^{-2}$, entropy in pockets of polar regions $\gtrsim15k_b\, \mathrm{baryon}^{-1}$, asymmetric shock), and thus can possibly lead to a successful explosion later on. Models \texttt{R10B12} and \texttt{R15B12} form a successful jet outflow, but the jet bends due to instabilities along the rotation axis, causing the entire shock front in the polar directions to tilt towards the equatorial plane. This effectively makes the shock front \textit{appear} roughly symmetric from an observational point of view, which means they could be attributed observationally to be \textit{symmetric} neutrino-driven explosions despite being powered by the magnetorotational mechanism. Model \texttt{R05B12} may appear observationally in a similar fashion. Finally, models \texttt{R20B12} and \texttt{R25B12} form a successful jet outflow moving steadily along the polar directions and being highly \textit{asymmetric}. 

\subsection{Shock evolution and propagation}

\begin{figure}
 	\includegraphics[width=0.45\textwidth]{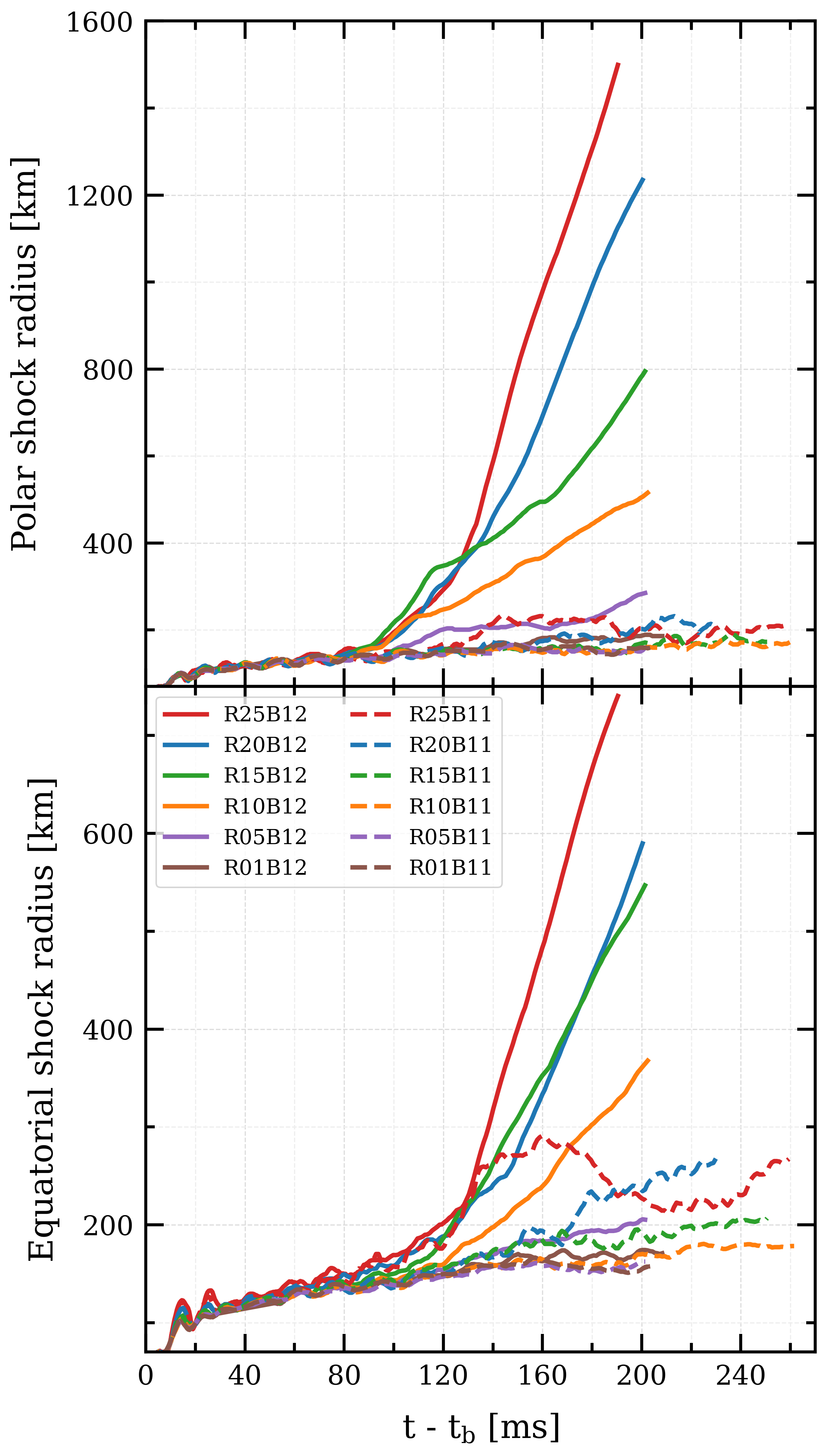}
\caption{\textit{Top panel:} Shock radius in the positive polar direction as a function of postbounce time for all models. \textit{Bottom panel:} Average shock radius in the equatorial direction as a function of postbounce time for all models. The \texttt{B12} models are shown with solid lines and the \texttt{B11} models are shown with dotted lines. Models with the same rotation rate are represented by the same color, e.g. solid red lines for model \texttt{R25B12} and dotted red lines for model \texttt{R25B11}. We note that the $y$-axis range is different for the top and bottom panels due to the shock expanding almost twice as far in the polar direction compared to the equatorial direction within the simulated time.} 
    \label{fig:shock_radius_vs_time}
\end{figure}

\begin{figure*}
 	\includegraphics[width=0.8\textwidth]{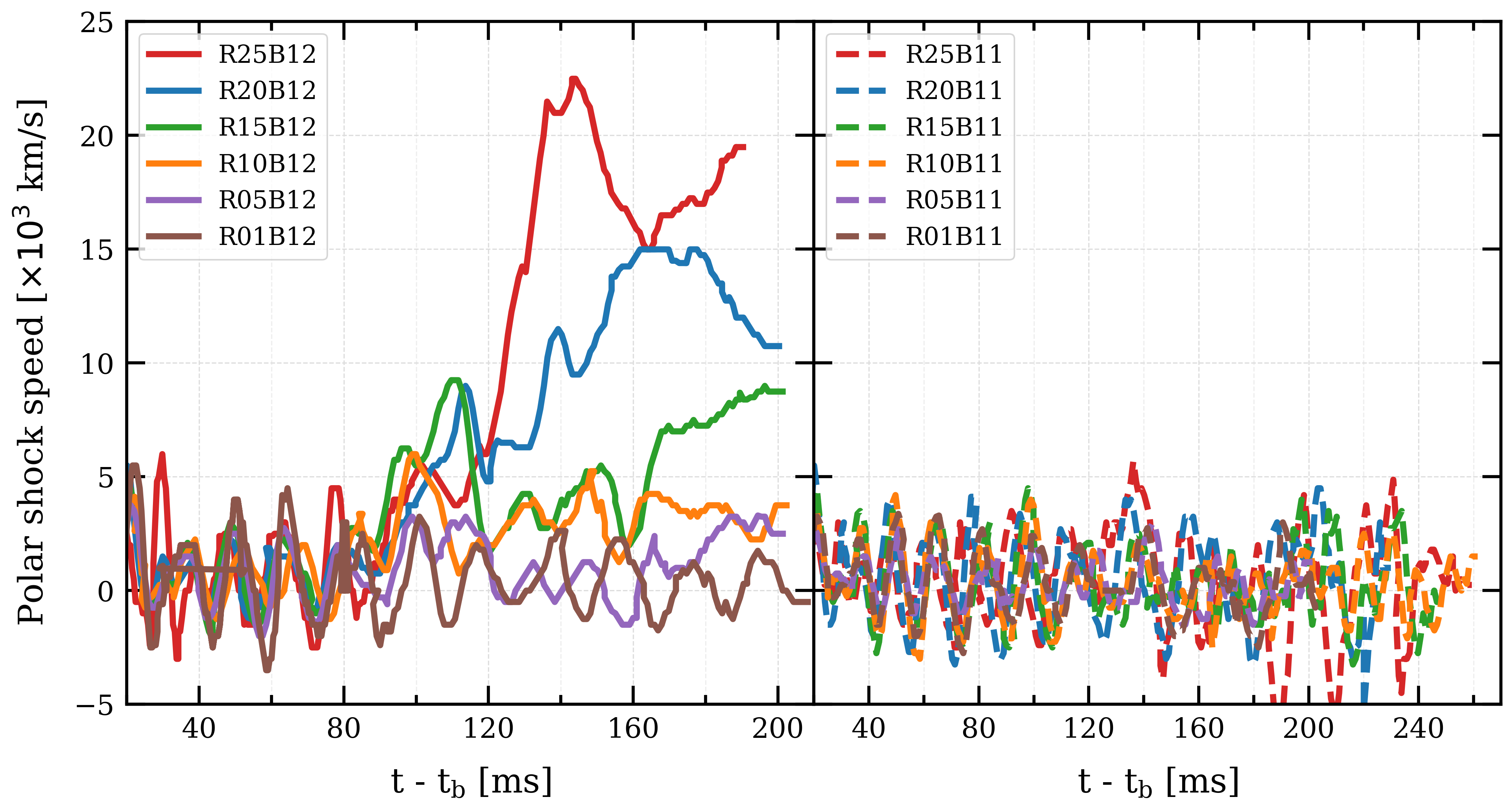}
    \caption{Shock propagation velocity in the polar direction as a function of postbounce time for all models. The left panel shows the shock velocity for the \texttt{B12} models with solid lines while the right panel shows the shock velocities for the \texttt{B11} models with dotted lines. We find that models \texttt{R20B12} and \texttt{R25B12} reach polar shock velocities $\gtrsim15000\, \mathrm{km/s}$ which is consistent with the shock velocities inferred observationally for broad-lines type-Ic supernovae~\citep{Modjaz_2016}. Model \texttt{R15B12} reaches polar shock velocities of $\sim10000\, \mathrm{km/s}$ which is at the upper limit of what is expected for traditional neutrino-driven core-collapse supernovae. All other models have polar shock velocities $\lesssim5000\, \mathrm{km/s}$ consistent with typical neutrino-driven core-collapse supernova.} 
    \label{fig:shock_speed_vs_time}
\end{figure*}

We show the shock radius in the positive $z$-direction in the left panel and the average equatorial shock radius in the right panel of Fig.~\ref{fig:shock_radius_vs_time}. We find that all models show very similar shock evolution at earlier times. Until $\sim80\, \mathrm{ms}$ postbounce, the shock remains nearly spherical with a radius that slowly increases from $\sim100\, \mathrm{km}$ at $20\, \mathrm{ms}$ to $\sim160\, \mathrm{km}$ at $80\, \mathrm{ms}$ while continuously sloshing around (i.e. oscillating) indicative of SASI~\citep{Iwakami_2009, Fernandez_2010, Endeve_2012}. After $80\, \mathrm{ms}$ however, the shock proceeds to evolve very differently between models. For all \texttt{B11} models the shock remains nearly spherical, with shock radii $\lesssim200\, \mathrm{km}$ along both the polar and the equatorial directions. Due to rotational flattening of the PNS and shock, the shock radii are slightly larger in the equatorial directions ($\sim300\, \mathrm{km}$) compared to the polar direction ($\sim200\, \mathrm{km}$) in the rapidly rotating models \texttt{R20B11} and \texttt{R25B11}. In the \texttt{B12} models, on the other hand, the shock radii show a wide range of outcomes with polar shock radii ranging from $\sim200\, \mathrm{km}$ in model \texttt{R01B12} to $\sim1500\, \mathrm{km}$ in model \texttt{R25B12}. The evolution of the shock radius in the polar direction is strongly correlated with the initial rotation rate $\Omega_0$ among the \texttt{B12} models, as expected. The shock propagates more rapidly in the polar direction for higher rotation rates as the outflow of highly-magnetized material from the proto-neutron star is more efficient. The shock evolution also differs between the positive and negative polar directions, as is evident from Fig.~\ref{fig:andes_set2}, but we only show the shock radii in the positive polar direction in Fig.~\ref{fig:shock_radius_vs_time} for clarity. The shock effectively remains stalled until the end for all \texttt{B11} models and for model \texttt{R01B12}. However, \texttt{B12} models with $\Omega_0 \geq 1.0\, \mathrm{rad/s}$ have steadily expanding shock waves in both polar and equatorial directions. Model \texttt{R05B12} showing signs of the shock stalling at $\sim160\, \mathrm{ms}$ postbounce, but the shock keeps slowly expanding further thereafter in both directions. Models \texttt{R20B12} and \texttt{R25B12} show rapid shock radius growth in the polar directions without showing any signs of saturation until the end of the simulations. This is in contrast with \cite{Shibagaki_2024} in which case even the most rapidly rotating model, \texttt{R20B12}, starts showing saturation in shock radius growth after $\gtrsim100\, \mathrm{ms}$ postbounce. This difference could be due to different input physics but maybe even more likely due to differences in resolution. The simulations of \cite{Shibagaki_2024} use a static AMR grid, and the resolution keeps decreasing as the shock expands further and moves to refinement levels with lower resolution while we keep the shock resolved at the same resolution and change our AMR grid setup as the shock expands.

We show the shock propagation velocity for the \texttt{B12} models in the left panel and the \texttt{B11} models in the right panel of Fig.~\ref{fig:shock_speed_vs_time}. We find that the shock propagation velocity for all models oscillates between positive and negative values (within $5000\, \mathrm{km/s}$) until $\sim80\, \mathrm{ms}$ postbounce for both polar and equatorial directions. For all \texttt{B11} models, this trend continues until the end as the shock wave does not successfully propagate out of the core of the star. The shock propagation velocity for the \texttt{B12} models shows a wide variety of outcomes, depending on the initial rotation rate $\Omega_0$. The shock velocity in models \texttt{R01B12}, \texttt{R05B12} and \texttt{R10B12} stays within  $\lesssim5000\, \mathrm{km/s}$ until the end, similar to the \texttt{B11} models. For model \texttt{R15B12} the shock propagates with velocity of almost $\sim10000\, \mathrm{km/s}$ at the end of the simulation, which is at the upper limit of the typical speeds in neutrino-driven CCSNe. Models \texttt{R20B12} and \texttt{R25B12} reach peak shock velocities between $\sim15000-25000\, \mathrm{km/s}$ which are characteristic of outflow velocities inferred for type Ic-bl CCSNe~\citep{Modjaz_2016}.  

\subsection{PNS evolution and accretion}

\begin{figure*}
 	\includegraphics[width=\columnwidth]{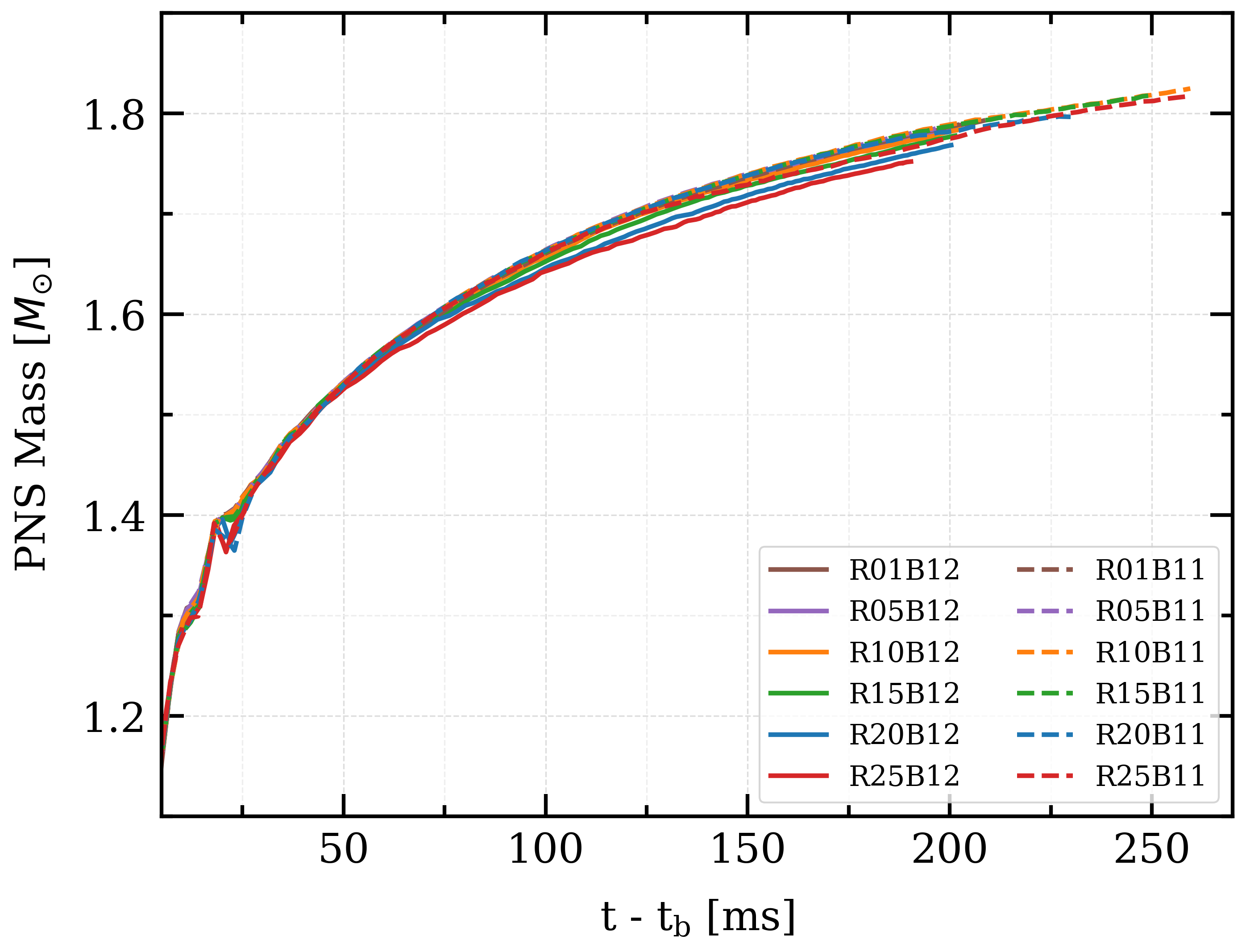}
        \includegraphics[width=\columnwidth]{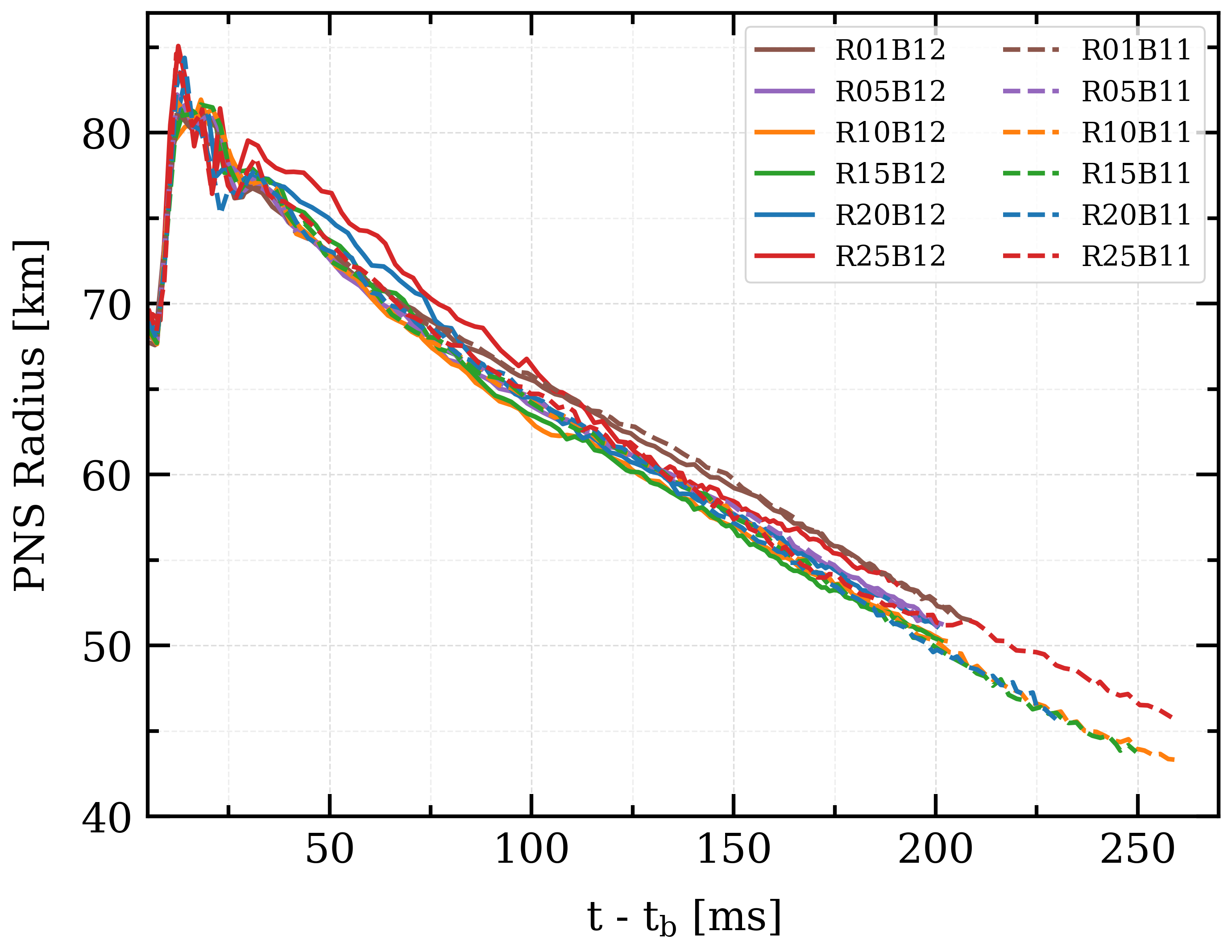}
\caption{\textit{Left panel}: Mass of the PNS as a function of postbounce time for all models. We find that the PNS mass keeps increasing continuously due to accretion and reaches $\sim1.8\,M_{\odot}$ towards the end of the simulated time for all models. The growth of the PNS mass shows asymptotic behavior at later time. 
\textit{Right panel}: Radius of the PNS as a function of postbounce time. The PNS radius keeps decreasing continuously from the peak value of $\sim85\, \mathrm{km}$ and reaches $\sim51-54\, \mathrm{km}$ at $190\, \mathrm{ms}$ postbounce for all models. } 
    \label{fig:PNS_mass_vs_time}
\end{figure*}

\begin{figure*}
 	\includegraphics[width=0.75\textwidth]{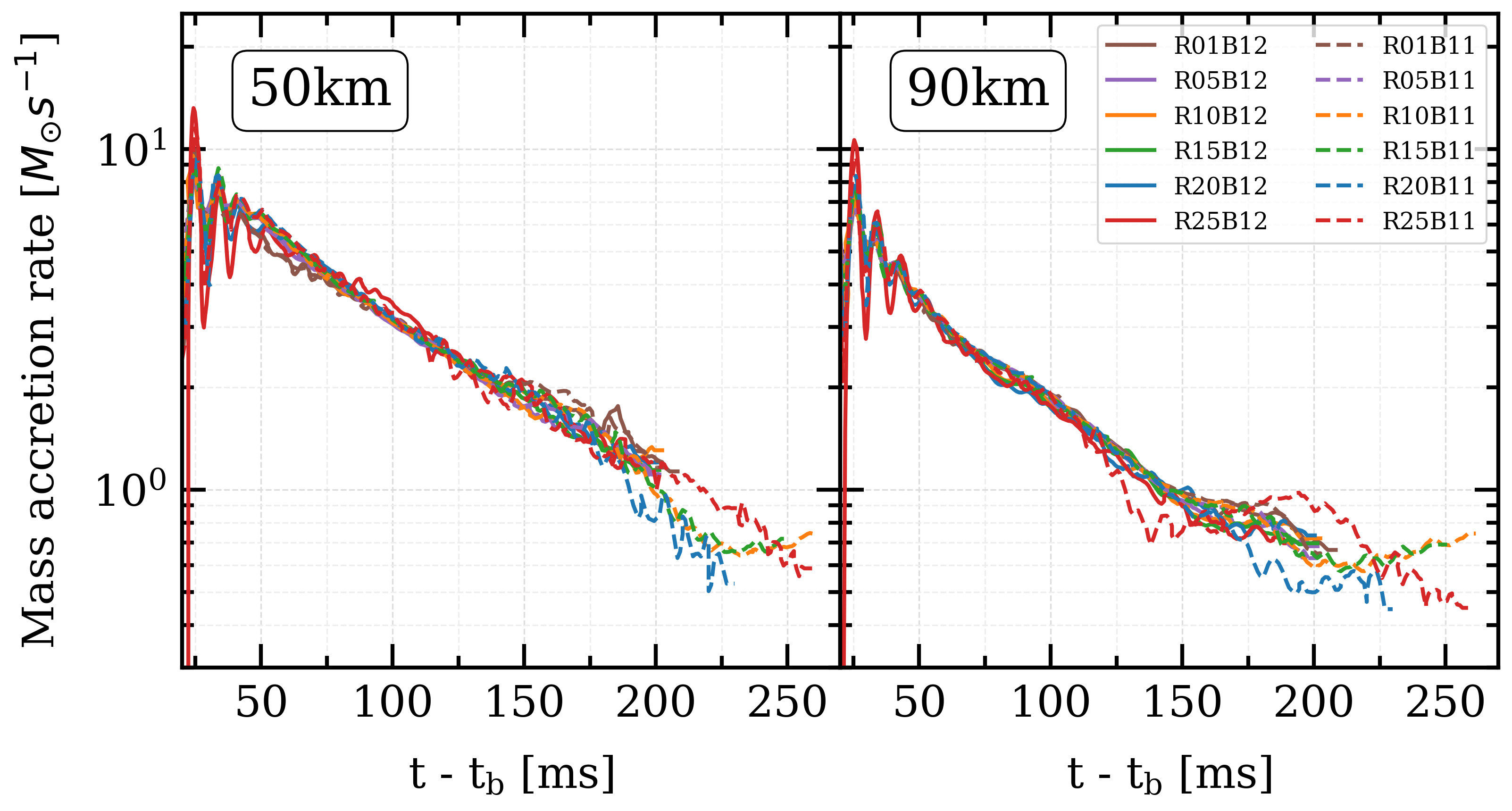}
\caption{Mass accretion rate for all models calculated at radii $50\, \mathrm{km}$ and $90\, \mathrm{km}$ in the left and the right panel, respectively. These radii are selected to bracket the evolving PNS surface, which shrinks from $\sim85\, \mathrm{km}$ at early times to $\sim50\, \mathrm{km}$ at $200\, \mathrm{ms}$, allowing us to probe the changing accretion conditions from near the initial PNS surface (90 km) to deeper layers closer to its contracting core (50 km). Solid lines show the accretion rate for the \texttt{B12} models, while dotted lines show the accretion rate for the \texttt{B11} models. We find that there is no significant model variability in the time evolution of accretion rate. It is effectively $\sim8\,M_{\odot}/\mathrm{s}$ at early times but keeps decreasing continuously and drops below $\sim1\,M_{\odot}/\mathrm{s}$ at $200\, \mathrm{ms}$ postbounce for all models. } 
    \label{fig:accretion_rate_vs_time}
\end{figure*}

We define the radius of the PNS ($R_{\mathrm{PNS}}$) as the isosurface with $\rho=10^{11}$~g~cm$^{-3}$ and calculate the mass of the PNS ($M_{\mathrm{PNS}}$) as the total mass enclosed within that surface. We show the mass and radius of the PNS for all models as a function of time in the left and right panels of Fig.~\ref{fig:PNS_mass_vs_time}. We find that the evolution of the PNS mass and PNS radius does not show any significant differences between the different models. The mass of the PNS increases monotonically due to accretion and reaches $\sim1.8\,M_{\odot}$ at $260\, \mathrm{ms}$ postbounce. Towards the end of the simulations, the PNS mass grows less fast as the accretion rate across the shock decreases. Our PNS masses are consistent with those obtained by \cite{Halevi_2023} for the 1D unmagnetized evolution of the same progenitor. The radius of the PNS is $\sim85\, \mathrm{km}$ initially and decreases monotonically towards a value of $\sim45\, \mathrm{km}$ at $260\, \mathrm{ms}$ postbounce as the PNS core becomes more dense with time due to accretion. The models show a variation of $\lesssim2\%$ in PNS masses with values in the range $1.75-1.78\, M_\odot$ at $\sim190\, \mathrm{ms}$ postbounce, and a variation of $\lesssim5\%$ in PNS radii with values in the range $51-54\, \mathrm{km}$ at $\sim190\, \mathrm{ms}$ postbounce. Less magnetized and slowly rotating models have slightly higher PNS masses. \cite{Shibagaki_2024} find a similar behavior of PNS masses for their $20\, M_\odot$ progenitor model wherein their model \texttt{R05B12} has a higher PNS mass ($\sim1.6\, M_\odot$) compared to their model \texttt{R20B12} ($\sim1.4\, M_\odot$), but in this case the rotation rate has a larger effect on the PNS mass, $\sim10\%$ as opposed to $\lesssim2\%$ seen in our models. \cite{Obergaulinger_2021} also find higher PNS masses for their less magnetized models, wherein the PNS mass varies from $\sim2.1\, M_\odot$ in model \texttt{W} ("Weak") to $\sim1.8\, M_\odot$ in model \texttt{S} ("Strong") which is a variation of $\sim15\%$. For PNS radii, all \texttt{B11} models have lower PNS radii compared to the \texttt{B12} models. Interestingly, models with $\Omega_0=1.5\, \mathrm{rad/s}$ show the lowest PNS radii at a given time, for both \texttt{B11} and \texttt{B12} models, with radii increasing when rotation rate is increased or decreased from $1.5\, \mathrm{rad/s}$ for a given $B_0$.

In order to study the accretion rate on the PNS quantitatively, we calculate the total mass enclosed within radii $50\, \mathrm{km}$ and $90\, \mathrm{km}$ as a function of time for all models. We then use the rate of change of this enclosed mass to calculate the effective accretion rate. We plot the accretion rate at $50\, \mathrm{km}$ and $90\, \mathrm{km}$ in Fig.~\ref{fig:accretion_rate_vs_time} and find that there are no significant differences between the accretion rates between the different models. The accretion rate is similar at $50\, \mathrm{km}$ and $90\, \mathrm{km}$ at early times ($\lesssim40\, \mathrm{ms}$ postbounce), with an effective value of $\sim8\,M_{\odot}/\mathrm{s}$. At later times, the accretion rate steadily decreases, with the rate being higher at $50\, \mathrm{km}$ compared to at $90\, \mathrm{km}$. The accretion rate at $50(90)\, \mathrm{km}$ drops to $\sim3(2)\,M_{\odot}/\mathrm{s}$ at $100\, \mathrm{ms}$ postbounce and $\sim2(1)\, M_{\odot}/\mathrm{s}$ at $150\, \mathrm{ms}$ postbounce. At $200\, \mathrm{ms}$ after bounce, the accretion rate has dropped below $\sim1\, M_{\odot}/\mathrm{s}$ for both radii, after which the rate appears to decrease even further to $\sim0.5-0.8\,M_{\odot}/\mathrm{s}$ for the different models. If accretion were to continue in this fashion, the PNS would form a black hole in $\approx1-2\, \mathrm{s}$ in our simulations.

\subsection{Explosion energy and ejecta properties}
\label{subsec:explosion_energy_ejecta_properties}
\begin{figure}
 	\includegraphics[width=\columnwidth]{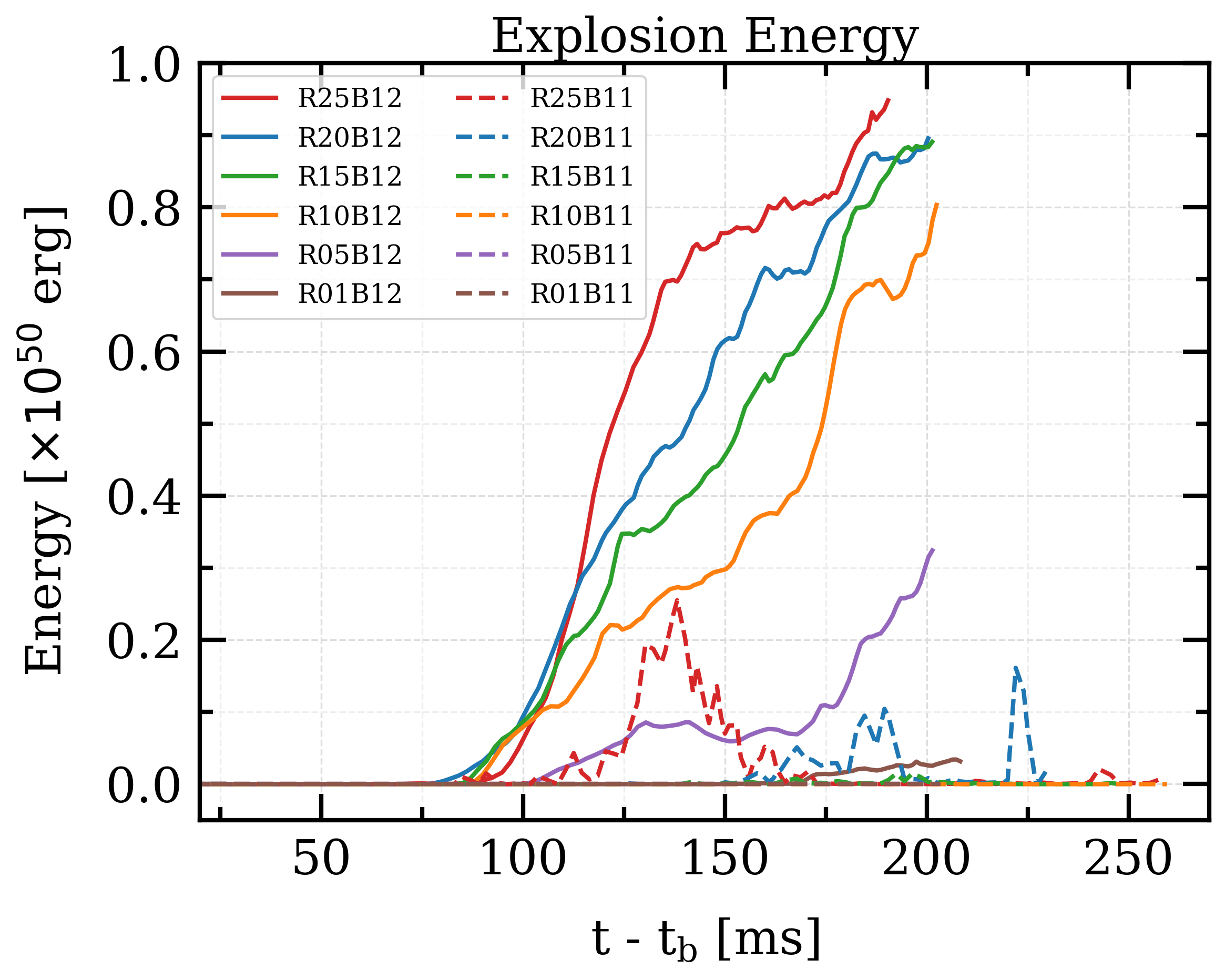}
        \includegraphics[width=\columnwidth]{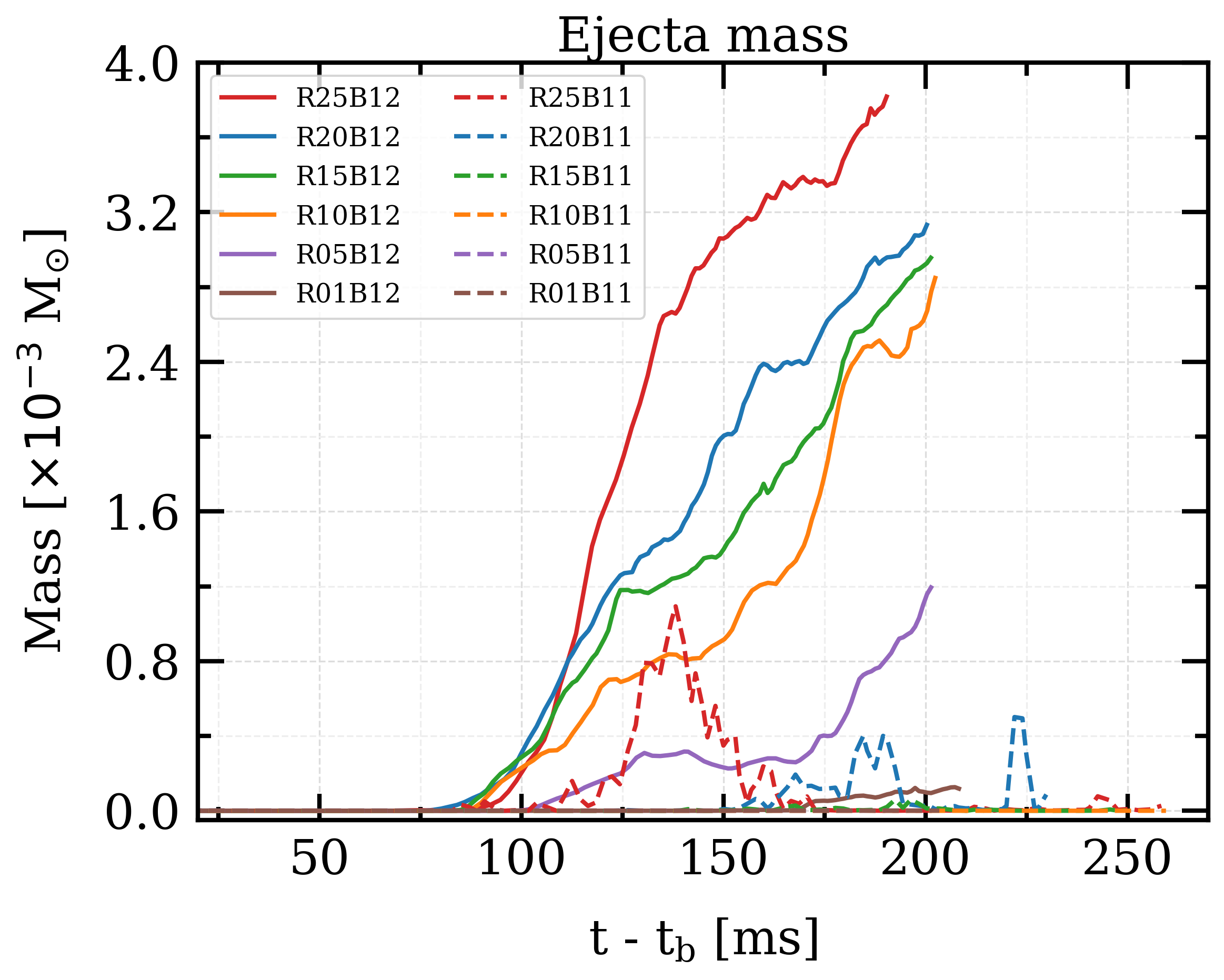}
    \caption{\textit{Top panel}: Explosion energy as a function of postbounce time for all models. We calculate the explosion energy as the sum of internal, kinetic and magnetic energy components as defined in Eq.~\ref{eq:E_kinetic}, \ref{eq:E_internal} and \ref{eq:E_magnetic} respectively. We find that the \texttt{B11} models, shown using dotted lines, have very  small explosion energies even at late times. The \texttt{B12} models, shown using solid lines,  show a wide range of explosion energies depending on the initial rotation rate $\Omega_0$. The explosion energies for the \texttt{B12} models keep increasing with time at varying rates and reach values of $\sim0.3-0.9\times10^{50}\, \mathrm{erg}$ for models with $\Omega_0\geq0.5\, \mathrm{rad/s}$ at the end of the simulated time. They are still increasing at the end of the simulations. \\  
\textit{Bottom panel}: Mass of the ejecta as a function of postbounce time for all models. We calculate the ejecta mass using Eq.~\ref{eq:M_ejecta}. Similar to the explosion energy, the \texttt{B11} models show very small ejecta masses even at late times, while the ejecta masses for the \texttt{B12} models keep increasing until the end of the simulated time, reaching values in the range $\sim1.2-4\times10^{-3} \, M_\odot$ for models with $\Omega_0\geq0.5\, \mathrm{rad/s}$.} 
    \label{fig:explosion_energy_vs_time}
\end{figure}

We use the Bernoulli criterion to determine which fluid elements are unbound. Under the Bernoulli criterion fluid elements are considered unbound if $-h\,u_t>1$ \citep[e.g.][]{Kastaun_2015}, where $h$ is the fluid specific enthalpy and $u_t$ is the covariant time component of the fluid 4-velocity. We calculate the diagnostic explosion energy as the sum of kinetic ($E_{\text{Kinetic}}$), internal ($E_{\text{Internal}}$) and magnetic energy ($E_{\text{Magnetic}}$) of all unbound fluid elements. We calculate the ejecta mass ($M_{\text{Ejecta}}$) and these energy contributions using the following equations~\citep{Cook_2025}:
\begin{eqnarray}
M_{\text{Ejecta}} &=&  \int_{\text{unbound}} \sqrt{\gamma} W \rho \, \mathrm{d}^3x, \label{eq:M_ejecta} \\
E_{\text{Kinetic}} &=& \int_{\text{unbound}} \sqrt{\gamma} (W-1) \rho  \, \mathrm{d}^3x, \label{eq:E_kinetic}\\  
E_{\text{Internal}} &=& \int_{\text{unbound}} \sqrt{\gamma} W \rho  \epsilon \, \mathrm{d}^3x, \label{eq:E_internal}\\
E_{\text{Magnetic}} &=& \frac{1}{2} \int_{\text{unbound}} \sqrt{\gamma} W b^2 \, \mathrm{d}^3x. \label{eq:E_magnetic}
\end{eqnarray}
where $\gamma$ is the determinant of the 3-metric, $W$ is the lorentz factor, $\rho$ is the fluid rest mass density, $\epsilon$ is the specific internal energy and $b^2=b_ib^i$ is the co-moving magnetic field. 

We show the explosion energy and ejecta mass as a function of time for all models in the top and bottom panels of Fig.~\ref{fig:explosion_energy_vs_time}. Explosion energies and ejecta masses are very small for all \texttt{B11} models, indicating that very little or no material gets unbound as these models do not explode. The \texttt{B12} models, on the other hand, show a range of explosion energies depending on the rotation rate, with more strongly rotating models showing higher energies, as expected.  The explosion energies for all \texttt{B12} models with  $\Omega_0\geq0.5\, \mathrm{rad/s}$ keep increasing steadily and do not show any signs of saturation throughout the simulated time. The rate of increase of explosion energy is not uniform and proceeds in "phases" of steep growth followed by slow growth. However, they only reach modest values of $0.3-0.9\times10^{50}\, \mathrm{erg}$ by $\sim200\, \mathrm{ms}$. These are much lower than the canonical value of $\sim10^{52}\, \mathrm{erg}$ expected for type Ic-bl CCSNe, however the explosion energies are expected to grow for a long time as material is constantly ejected from the PNS. The explosion energy in the simulations of \cite{Shibagaki_2024} reaches $\sim1.0\times10^{50}\, \mathrm{erg}$ for their model \texttt{R10B12} and $\sim6.0\times10^{50}\, \mathrm{erg}$ for their model \texttt{R20B12} at $200\, \mathrm{ms}$ postbounce, which is higher than our models but still much lower than $10^{52}\, \mathrm{erg}$. In their models, the explosion energy saturates and even starts decreasing after $300\, \mathrm{ms}$ postbounce, but this could possibly be due to the shock being poorly resolved at late times since the shock reaches regions with a resolution of $\gtrsim100\, \mathrm{km}$. Explosion energies of a few times $10^{50}\, \mathrm{erg}$ after $\sim300-500\, \mathrm{ms}$ postbounce are also seen in the 3D models of \cite{Bugli_2021} and \cite{Powell_2023}. The explosion energy of the strongly-magnetized model \texttt{S} of \cite{Obergaulinger_2021} reaches $1.3\times10^{52}\, \mathrm{erg}$ at $1.15\, \mathrm{s}$ postbounce and still keeps slowly increasing at that time. For neutrino-driven explosions, the explosion energy can take $\gtrsim2-3\, \mathrm{s}$ to saturate, e.g. the explosion energy is still slowly increasing after $2\, \mathrm{s}$ in the neutrino-driven simulations of \cite{Muller_2017}.  In general, our explosion energies at $\sim200\, \mathrm{ms}$ postbounce are smaller compared to other 3D studies for similar magnetic field strengths but these relatively low explosion energies can be attributed to the high compactness ($\xi_{2.5} = 0.47$) of the progenitor (it does not explode in 1D simulations carried out without magnetic fields \citep{Halevi_2023}).

We show the ejecta mass for all models in the bottom panel of Fig.~\ref{fig:explosion_energy_vs_time}. We calculate the ejecta mass as the sum of the mass of all the fluid elements which are unbound. As expected, the ejecta mass shows similar behavior as the explosion energy for all models. The ejecta mass is negligible for all \texttt{B11} models. For the \texttt{B12} models, the ejecta mass is higher for rapidly rotating models, as expected. At the end of the simulations, it reaches a value of $\sim2.8-4\times10^{-3}\, M_\odot$ for models \texttt{R10B12} to \texttt{R25B12},  $\sim1.2\times10^{-3}\, M_\odot$ for model \texttt{R05B12} while being negligibly small for model \texttt{R01B12}. As is the case for the explosion energy, the ejecta mass does not saturate until the end of the simulated times, increasing at varying rates throughout the simulations for models with  $\Omega_0\geq0.5\, \mathrm{rad/s}$. 


\begin{table*}
    \centering
    \begin{tabular}{lccccccccccc}
        \hline
        Model & $t_b$ & $t_{\mathrm{end}}$ & $M_{\mathrm{PNS}}$ & $R_{\mathrm{PNS}}$ & $\dot{M}_{50\mathrm{km}}$ & $R^{\mathrm{shock}}_{\mathrm{po,avg}}$ & $R^{\mathrm{shock}}_{\mathrm{eq,avg}}$ & $v^{\mathrm{shock}}_{\mathrm{po, max}}$ & $M_{\mathrm{ej}}$ & $E_{\mathrm{exp}}$ & $Z$\\
         & [ms] & [ms] & [$M_\odot$] & [km] & [$M_\odot/s$] & [km] & [km] & [$\mathrm{km/s}$] & [$M_\odot$] & [erg] & \\
        \hline
        \texttt{R01B11} & 163.7 & 204 & 1.79 & 52.5 & $1.16\times 10^{-3}$ & 157 & 155 & 5400 & 0.0 & 0.0 & 0.01 \\
        \hline
        \texttt{R05B11} & 163.8 & 202 & 1.79 & 51.0 & $1.11\times 10^{-3}$ & 153 & 162 & 5000 & 0.0 & 0.0 & 0.03\\
        \hline
        \texttt{R10B11} & 164.1 & 261 & 1.79 & 50.0 & $0.96\times 10^{-3}$ & 160 & 170 & 5000 & $1.12\times 10^{-8}$ &  $2.78\times10^{44}$ & 0.03\\
        \hline
        \texttt{R15B11} & 164.7 & 251 & 1.79 & 49.8 & $1.00\times 10^{-3}$ & 169 & 189 & 6000 & $1.18\times 10^{-5}$ & $2.99\times10^{47}$ & 0.06 \\
        \hline
        \texttt{R20B11} & 165.6 & 231 & 1.78 & 49.7 & $0.82\times 10^{-3}$ & 194 & 234 & 5500 & $3.37\times 10^{-5}$ & $8.31\times10^{47}$ & 0.09  \\
        \hline
        \texttt{R25B11} & 166.6 & 260 & 1.77 & 51.3 & $0.99\times 10^{-3}$ & 201 & 227 & 5700 & $3.29\times 10^{-7}$ & $1.06\times10^{46}$ & 0.06 \\
        \hline
        \texttt{R01B12} & 163.7 & 209 & 1.79 & 52.3 & $1.22\times 10^{-3}$ & 185 & 174 & 6600 & $1.01\times 10^{-4}$ & $2.70\times10^{48}$ & 0.03\\
        \hline
        \texttt{R05B12} & 163.8 & 202 & 1.78 & 51.3 & $1.11\times 10^{-3}$ & 280 & 206 & 5700 & $1.16\times 10^{-3}$ & $3.15\times10^{49}$ & 0.15\\
        \hline
        \texttt{R10B12} & 164.2 & 203 & 1.78 & 50.5 & $1.30\times 10^{-3}$ & 532 & 362 & 6000 & $2.67\times 10^{-3}$ & $7.51\times10^{49}$ & 0.19 \\
        \hline
        \texttt{R15B12} & 164.8 & 202 & 1.78 & 50.4 & $1.14\times 10^{-3}$ & 750 & 547 & 9200 & $2.92\times 10^{-3}$ & $8.84\times10^{49}$ & 0.16\\
        \hline
        \texttt{R20B12} & 165.6 & 200 & 1.77 & 51.1 & $1.20\times 10^{-3}$ & 1136 & 590 & 15000 & $3.13\times 10^{-3}$ & $8.96\times10^{49}$ & 0.32 \\
        \hline
        \texttt{R25B12} & 166.7 & 191 & 1.75~$^\dagger$ & 53.6~$^\dagger$ & $1.23\times 10^{-3}$~$^\dagger$ & 1364~$^\dagger$ & 740~$^\dagger$ & 22500 & $3.82\times 10^{-3}$~$^\dagger$ & $9.48\times10^{49}$~$^\dagger$ & 0.30 \\
        \hline
    \end{tabular} \\
    \footnotesize{$^\dagger$ Quantities for model \texttt{R25B12} displayed at $190\, \mathrm{ms}$ postbounce. All other quantities displayed at $200\, \mathrm{ms}$ postbounce. }
    \caption{Summary of key results for all simulated models at $200\, \mathrm{ms}$ postbounce ($190\, \mathrm{ms}$ postbounce for model \texttt{R25B12}). From left to right, the columns show the model name, time of bounce ($t_b$) in ms, end time of the simulation ($t_{\mathrm{end}}$) in ms, PNS mass ($M_{\mathrm{PNS}}$) in $M_\odot$, PNS radius ($R_{\mathrm{PNS}}$) in km, mass accretion rate at $50\, \mathrm{km}$ ($\dot{M}_{50\mathrm{km}}$) in $M_\odot/s$, average polar shock radius ($R^{\mathrm{shock}}_{\mathrm{po,avg}}$) in km, average equatorial shock radius ($R^{\mathrm{shock}}_{\mathrm{eq,avg}}$) in km,  maximum shock velocity in the polar direction ($v^{\mathrm{shock}}_{\mathrm{po, max}}$) in $\mathrm{km/s}$, mass of the ejecta ($M_{\mathrm{ej}}$) in $M_\odot$, explosion energy ($E_{\mathrm{exp}}$) in erg, and the asphericity parameter $Z$ as defined in eq.~\ref{eq:definition_Z}. }
    \label{tab:all_simulation_summary}
    
\end{table*}

We summarize key results for all models in Table~\ref{tab:all_simulation_summary} at $\sim200\, \mathrm{ms}$ postbounce ($\sim190\, \mathrm{ms}$ postbounce for model \texttt{R25B12} as the simulation becomes very expensive at later times due to the fast expanding shock). We list bounce time $\equiv t_b$, end time of the simulation $\equiv t_{\mathrm{end}}$, mass of the PNS $\equiv M_{\mathrm{PNS}}$, radius of the PNS $\equiv R_{\mathrm{PNS}}$, mass accretion rate at $50\, \mathrm{km} \equiv \dot{M}_{50\mathrm{km}}$, average shock radius in the polar direction $\equiv R^{\mathrm{shock}}_{\mathrm{po,avg}}$, average shock radius in the equatorial direction $\equiv R^{\mathrm{shock}}_{\mathrm{eq,avg}}$,  maximum shock speed in the polar direction $\equiv v^{\mathrm{shock}}_{\mathrm{po, max}}$, mass of the ejecta $\equiv M_{\mathrm{ej}}$ and explosion energy $\equiv E_{\mathrm{exp}}$. We define the asphericity parameter, $Z$, as follows
\begin{equation}
\label{eq:definition_Z}
Z = \frac{|R^{\mathrm{shock}}_{\mathrm{po,avg}} - R^{\mathrm{shock}}_{\mathrm{eq,avg}}|}{R^{\mathrm{shock}}_{\mathrm{po,avg}} +R^{\mathrm{shock}}_{\mathrm{eq,avg}}}  
\end{equation}
such that $Z \rightarrow 0$ for a spherical shock profile and $Z \rightarrow 1$ when shock is highly aspherical i.e $R^{\mathrm{shock}}_{\mathrm{po,avg}} \gg R^{\mathrm{shock}}_{\mathrm{eq,avg}}$ or $R^{\mathrm{shock}}_{\mathrm{eq,avg}} \gg R^{\mathrm{shock}}_{\mathrm{po,avg}}$. Thus, $0 < Z < 1$ and higher values of $Z$ indicate higher asphericity. We list $Z$ for all the models in the last column of Table~\ref{tab:all_simulation_summary}. We find that models \texttt{R10B12} and \texttt{R15B12} show similar asphericity as model \texttt{R05B12} despite having much more favorable jet formation parameters, because the formed bipolar outflow tilts towards the equatorial plane in both \texttt{R10B12} and \texttt{R15B12}. The shock \textit{appears} spherical despite being jet-driven, especially for model \texttt{R15B12}.

\section{Conclusions}
\label{sec:conclusions}
Rotation and magnetic fields play a crucial role in the magnetorotational CCSNe mechanism. However, the rotation profiles and magnetic fields of massive stars at the end of their lives are largely unconstrained because of a lack of observational data and the fact that 1D stellar evolution simulations cannot fully capture the relevant physics and timescales at play. This necessitates systematic studies of the effects of rotation and magnetic fields to explore conditions for explodability and explosion parameters. This has traditionally been limited due to the need of doing these simulations in 3D to capture the relevant MHD processes driving the explosion and thus result in large computational costs. In this work, we present a series of three-dimensional dynamical-spacetime GRMHD simulations of a progenitor with zero-age-main-sequence mass of $25\, M_\odot$ and systematically study the effect of rotation rates and magnetic fields on jet formation via the magnetorotational mechanism. We simulate 12 models using magnetic field strengths, $B_0 = (10^{11}, 10^{12})\, \mathrm{G}$, and rotation rates, $\Omega_0 = (0.14, 0.5, 1.0, 1.5, 2.0, 2.5)\, \mathrm{rad/s}$, with the GPU-accelerated ideal-GRMHD code \texttt{GRaM-X}. 


We find that all models with $B_0 = 10^{11}\, \mathrm{G}$ fail to launch a successful jet and do not explode until the end of the simulations, regardless of the initial rotation rate $\Omega_0$. On the other hand, models with $B_0 = 10^{12}\, \mathrm{G}$ show a wide range of jet morphologies and explosion properties depending on $\Omega_0$. We characterized the resulting explosion geometries to set the stage for connecting these simulations to observations. Models with $B_0 = 10^{12}\, \mathrm{G}$ and $\Omega_0 \leq 0.5\, \mathrm{rad/s}$ fail to launch a successful jet and the shock remains generally spherically symmetric until the end of the simulations. Models with $B_0 = 10^{12}\, \mathrm{G}$ and $\Omega_0 = (1.0, 1.5)\, \mathrm{rad/s}$ are able to launch a bipolar jet and successfully explode, but the jet gets distorted by MHD plasma instabilities and a portion of it falls back onto the PNS along the equatorial direction. As a result, these explosions do not have strong directionality and \textit{appear} generally spherically symmetric. This implies, that their explosion geometry and energy appears similar to neutrino-driven CCSNe despite being powered by the magnetorotational mechanism. Models with $B_0 = 10^{12}\, \mathrm{G}$ and $\Omega_0\geq 2.0\, \mathrm{rad/s}$ launch successful jets which move steadily outwards in the polar direction with velocities $\gtrsim15000\, \mathrm{km/s}$, making them suitable candidates for broad-lined type Ic supernovae. We find that the PNS masses and accretion rates are similar across all models, while the shock radii and explosion energies show a wide range of outcomes depending on $\Omega_0$ and $B_0$. For all models that do not explode, it is possible that a later explosion occurs either due to neutrino heating or due to a delayed buildup of strong, large-scale magnetic fields.

To the best of our knowledge, this work represents the largest set of 3D GRMHD core-collapse supernova models employing the magnetorotational mechanism in dynamical GR. With this, we demonstrate the potential in GPU-accelerated simulations for systematic parameter studies of MHD-powered CCSNe in 3D. The total cost of simulating all models presented here was $~\sim35000$~GPU node-hours on OLCF's Frontier, less than 10\% of a typical INCITE/tier-0 allocation for a given year. We have used the computationally less expensive M0 neutrino-transport approximation to be able to afford high resolution throughout the entire shocked region. This ensures that we always cover the entire region containing the shock with resolution of at least 1.48 km. We find that our exploding models show steady shock expansion without any signs of saturation until the end ($\sim200\mathrm{ms}$). This is a key difference to studies that have employed more sophisticated neutrino transport but do not resolve the shocked region with constant high resolution and find that, for example, the shock radius starts to saturate after $\sim100\, \mathrm{ms}$ post-bounce for even the strongest exploding models \citep{Shibagaki_2024}. This result highlights the importance of resolution in the magnetorotational CCSN simulations.

We have used the less computationally expensive M0 neutrino transport in this work since our focus was on explosive outcome and neutrinos being energetically sub-dominant in magnetorotational CCSNe. We are currently performing follow-up simulations of select models using the M1 neutrino transport to understand differences arising out of the neutrino treatment and to study the nucleosynthetic properties of the ejected material. There is also a clear need for long-term ($\gtrsim1-2\, \mathrm{s}$) simulations of these models in order to understand the fate of the explosion and determine the central remnant left behind. In this work, we have only considered a single $25\, M_\odot$ high-compactness progenitor to focus on the impact of rotation and magnetic fields, but in the future we plan to explore progenitors of lower compactness and different masses. Type Ic-bl SNe show ejecta velocities as high as $30000\, \mathrm{km/s}$, but in the current work the strongest models reach peak velocities in the range $15000-25000\, \mathrm{km/s}$, which we largely attribute to the high-compactness progenitor we have chosen. We expect a lower compactness progenitor to reach higher velocities, possibly reaching $30000\, \mathrm{km/s}$. The explosion energy at the end of the simulation time for the strongest models is also $\sim10^{50}\, \mathrm{erg}$ which is well below the canonical type Ic-bl SN energy of $\sim10^{52}\, \mathrm{erg}$, but this is to be expected at these early times as the explosion energy can typically only be reliably determined at later times \citep[$\gtrsim2-3\, \mathrm{s}$;][]{Muller_2017}. 

\section*{Acknowledgments}

SS would like to thank Sherwood Richers for hosting him as a postdoctoral fellow at the University of Tennessee in Knoxville where some of the work was carried out.
This work has benefited from participation in the GRaM-X hackathon 2024 at the Perimeter Institute for Theoretical Physics.
PM acknowledges funding through NWO under grant No. OCENW.XL21.XL21.038.
RH acknowledges support by NSF awards OAC-2004879, OAC-2005572, OAC-2103680,  OAC-2310548, OAC-2411068. 
ES and RH acknowledge the support of the Natural Sciences and Engineering
Research Council of Canada (NSERC).
Research at Perimeter Institute is supported in part by the Government
of Canada through the Department of Innovation, Science and Economic
Development and by the Province of Ontario through the Ministry of
Colleges and Universities.
This research used resources of the Oak Ridge Leadership Computing Facility at the Oak Ridge National Laboratory, which is supported by the Office of Science of the U.S. Department of Energy under Contract No. DE-AC05-00OR22725. The simulations were carried out on OLCF’s Frontier using the INCITE-2024 award under allocation AST191.

\section*{Data Availability}
The data underlying this article will be shared on reasonable request to the corresponding author.



\bibliographystyle{mnras}

\begin{thebibliography}{}
\makeatletter
\relax
\def\mn@urlcharsother{\let\do\@makeother \do\$\do\&\do\#\do\^\do\_\do\%\do\~}
\def\mn@doi{\begingroup\mn@urlcharsother \@ifnextchar [ {\mn@doi@}
  {\mn@doi@[]}}
\def\mn@doi@[#1]#2{\def\@tempa{#1}\ifx\@tempa\@empty \href
  {http://dx.doi.org/#2} {doi:#2}\else \href {http://dx.doi.org/#2} {#1}\fi
  \endgroup}
\def\mn@eprint#1#2{\mn@eprint@#1:#2::\@nil}
\def\mn@eprint@arXiv#1{\href {http://arxiv.org/abs/#1} {{\tt arXiv:#1}}}
\def\mn@eprint@dblp#1{\href {http://dblp.uni-trier.de/rec/bibtex/#1.xml}
  {dblp:#1}}
\def\mn@eprint@#1:#2:#3:#4\@nil{\def\@tempa {#1}\def\@tempb {#2}\def\@tempc
  {#3}\ifx \@tempc \@empty \let \@tempc \@tempb \let \@tempb \@tempa \fi \ifx
  \@tempb \@empty \def\@tempb {arXiv}\fi \@ifundefined
  {mn@eprint@\@tempb}{\@tempb:\@tempc}{\expandafter \expandafter \csname
  mn@eprint@\@tempb\endcsname \expandafter{\@tempc}}}

\bibitem[\protect\citeauthoryear{Aguilera-Dena, Langer, Antoniadis, Pauli,
  Dessart, Vigna-GÃ³mez, Gr\"afener  \& Yoon}{Aguilera-Dena
  et~al.}{2022}]{Aguilera-Dena_2020}
Aguilera-Dena D.~R.,  Langer N.,  Antoniadis J.,  Pauli D.,  Dessart L.,
  Vigna-Gomez A.,  Gr\"afener G.,   Yoon S.-C.,  2022, \mn@doi [A&A]
  {10.1051/0004-6361/202142895}, 661, A60

\bibitem[\protect\citeauthoryear{Akiyama, Wheeler, Meier  \&
  Lichtenstadt}{Akiyama et~al.}{2003}]{Akiyama2003}
Akiyama S.,  Wheeler J.~C.,  Meier D.~L.,   Lichtenstadt I.,  2003, \mn@doi
  [The Astrophysical Journal] {10.1086/344135}, 584, 954

\bibitem[\protect\citeauthoryear{{Balbus} \& {Hawley}}{{Balbus} \&
  {Hawley}}{1991}]{Balbus1991}
{Balbus} S.~A.,  {Hawley} J.~F.,  1991, \mn@doi [\apj] {10.1086/170270}, \href
  {https://ui.adsabs.harvard.edu/abs/1991ApJ...376..214B} {376, 214}

\bibitem[\protect\citeauthoryear{Banyuls, Font, Ibanez, Marti  \&
  Miralles}{Banyuls et~al.}{1997}]{Banyuls:1997zz}
Banyuls F.,  Font J.~A.,  Ibanez J.~M.,  Marti J.~M.,   Miralles J.~A.,  1997,
  Astrophys. J., 476, 221

\bibitem[\protect\citeauthoryear{Bugli, Guilet  \& Obergaulinger}{Bugli
  et~al.}{2021}]{Bugli_2021}
Bugli M.,  Guilet J.,   Obergaulinger M.,  2021, \mn@doi [Monthly Notices of
  the Royal Astronomical Society] {10.1093/mnras/stab2161}, 507, 443

\bibitem[\protect\citeauthoryear{Burrows \& Vartanyan}{Burrows \&
  Vartanyan}{2021}]{Burrows2021}
Burrows A.,  Vartanyan D.,  2021, \mn@doi [Nature]
  {10.1038/s41586-020-03059-w}, 589, 29

\bibitem[\protect\citeauthoryear{Burrows, Dessart, Livne, Ott  \&
  Murphy}{Burrows et~al.}{2007}]{Burrows_2007}
Burrows A.,  Dessart L.,  Livne E.,  Ott C.~D.,   Murphy J.,  2007, \mn@doi
  [The Astrophysical Journal] {10.1086/519161}, 664, 416

\bibitem[\protect\citeauthoryear{Cano, Wang, Dai  \& Wu}{Cano
  et~al.}{2017}]{Cano2017}
Cano Z.,  Wang S.-Q.,  Dai Z.-G.,   Wu X.-F.,  2017, \mn@doi [Advances in
  Astronomy] {10.1155/2017/8929054}, 2017, 8929054

\bibitem[\protect\citeauthoryear{{Cerd\'a-Dur\'an, P.}, {Font, J. A.},
  {Ant\'on, L.}  \& {M\"uller, E.}}{{Cerd\'a-Dur\'an, P.}
  et~al.}{2008}]{cerda_duran:2008}
{Cerd\'a-Dur\'an, P.} {Font, J. A.} {Ant\'on, L.}  {M\"uller, E.} 2008, \mn@doi
  [A\&A] {10.1051/0004-6361:200810086}, 492, 937

\bibitem[\protect\citeauthoryear{Cook, Daszuta, Fields, Hammond, Albanesi,
  Zappa, Bernuzzi  \& Radice}{Cook et~al.}{2025}]{Cook_2025}
Cook W.,  Daszuta B.,  Fields J.,  Hammond P.,  Albanesi S.,  Zappa F.,
  Bernuzzi S.,   Radice D.,  2025, \mn@doi [The Astrophysical Journal
  Supplement Series] {10.3847/1538-4365/ad87d4}, 277, 3

\bibitem[\protect\citeauthoryear{Curtis, Bosch, M\"osta, Radice, Bernuzzi,
  Perego, Haas  \& Schnetter}{Curtis et~al.}{2023}]{curtis_2023}
Curtis S.,  Bosch P.,  M\"osta P.,  Radice D.,  Bernuzzi S.,  Perego A.,  Haas
  R.,   Schnetter E.,  2023, Outflows from Short-Lived Neutron-Star Merger
  Remnants Can Produce a Blue Kilonova, 2023, \mn@doi[The Astrophysical 
  Journal Letters] {10.3847/2041-8213/ad0fe1}, 961, 1

\bibitem[\protect\citeauthoryear{Einfeldt}{Einfeldt}{1988}]{Einfeldt:1988og}
Einfeldt B.,  1988, \mn@doi [SIAM J. Numer. Anal.] {10.1137/0725021}, 25, 294

\bibitem[\protect\citeauthoryear{Endeve, Cardall, Budiardja, Beck, Bejnood,
  Toedte, Mezzacappa  \& Blondin}{Endeve et~al.}{2012}]{Endeve_2012}
Endeve E.,  Cardall C.~Y.,  Budiardja R.~D.,  Beck S.~W.,  Bejnood A.,  Toedte
  R.~J.,  Mezzacappa A.,   Blondin J.~M.,  2012, \mn@doi [The Astrophysical
  Journal] {10.1088/0004-637X/751/1/26}, 751, 26

\bibitem[\protect\citeauthoryear{Fernandez}{Fernandez}{2010}]{Fernandez_2010}
Fernandez R.,  2010, \mn@doi [The Astrophysical Journal]
  {10.1088/0004-637X/725/2/1563}, 725, 1563

\bibitem[\protect\citeauthoryear{Font}{Font}{2008}]{Font:2007zz}
Font J.~A.,  2008, Living Rev. Relativity, 11

\bibitem[\protect\citeauthoryear{Gammie, McKinney  \& Toth}{Gammie
  et~al.}{2003}]{Gammie:2003rj}
Gammie C.~F.,  McKinney J.~C.,   Toth G.,  2003, \mn@doi [Astrophys. J.]
  {10.1086/374594}, 589, 444

\bibitem[\protect\citeauthoryear{Haas et~al.,}{Haas
  et~al.}{2024}]{roland_haas_2024_zenodo}
Haas R.,  et~al., 2024, The Einstein Toolkit, \mn@doi{10.5281/zenodo.14193969},
  \url {https://doi.org/10.5281/zenodo.14193969}

\bibitem[\protect\citeauthoryear{Halevi \& M\"osta}{Halevi \&  
  M\"osta}{2018}]{Halevi_2018} 
Halevi G.,  M\"osta P.,  2018, \mn@doi [Monthly Notices of the Royal
  Astronomical Society] {10.1093/mnras/sty797}, 477, 2366

\bibitem[\protect\citeauthoryear{Halevi, Wu, M\"osta, Gottlieb, Tchekhovskoy  \&
  Aguilera-Dena}{Halevi et~al.}{2023}]{Halevi_2023}
Halevi G.,  Wu B.,  M\"osta P.,  Gottlieb O.,  Tchekhovskoy A.,   Aguilera-Dena
  D.~R.,  2023, \mn@doi [The Astrophysical Journal Letters]
  {10.3847/2041-8213/acb702}, 944, L38

\bibitem[\protect\citeauthoryear{{Harten}}{{Harten}}{1983}]{Harten:1983hr}
{Harten} A.,  1983, \mn@doi [J. Comp. Phys.] {10.1016/0021-9991(83)90136-5},
  \href {http://adsabs.harvard.edu/abs/1983JCoPh..49..357H} {49, 357}

\bibitem[\protect\citeauthoryear{Hilditch, Bernuzzi, Thierfelder, Cao, Tichy
  \& Br{\"u}gmann}{Hilditch et~al.}{2013}]{hilditch2013}
Hilditch D.,  Bernuzzi S.,  Thierfelder M.,  Cao Z.,  Tichy W.,   Br{\"u}gmann
  B.,  2013, \mn@doi [Phys. Rev. D] {10.1103/PhysRevD.88.084057}, 88, 084057

\bibitem[\protect\citeauthoryear{Hjorth \& Bloom}{Hjorth \&
  Bloom}{2012}]{hjorth_bloom_2012}
Hjorth J.,  Bloom J.~S.,  2012, The GRB-supernova connection.
Cambridge University Press, p. 169-190, \mn@doi{10.1017/CBO9780511980336.010}

\bibitem[\protect\citeauthoryear{Ib{\'a}{\~n}ez, Aloy, Font, Mart{\'\i},
  Miralles  \& Pons}{Ib{\'a}{\~n}ez et~al.}{1999}]{Ibanez:2001:godunov}
Ib{\'a}{\~n}ez J.,  Aloy M.,  Font J.,  Mart{\'\i} J.,  Miralles J.,   Pons J.,
   1999, in Toro E.,  ed., Proceedings of Godunov Methods, Theory and
  Applications 1999. Kluwer Academic/Plenum Publishers, New York, p.~485
  (\mn@eprint {} {arXiv:astro-ph/9911034})

\bibitem[\protect\citeauthoryear{Iwakami, Kotake, Ohnishi, Yamada  \&
  Sawada}{Iwakami et~al.}{2009}]{Iwakami_2009}
Iwakami W.,  Kotake K.,  Ohnishi N.,  Yamada S.,   Sawada K.,  2009, \mn@doi
  [The Astrophysical Journal] {10.1088/0004-637X/700/1/232}, 700, 232

\bibitem[\protect\citeauthoryear{Kalinani et~al.,}{Kalinani
  et~al.}{2024}]{Kalinani_2025}
Kalinani J.~V.,  et~al., 2024, \mn@doi [Classical and Quantum Gravity]
  {10.1088/1361-6382/ad9c11}, 42, 025016

\bibitem[\protect\citeauthoryear{Kastaun \& Galeazzi}{Kastaun \&
  Galeazzi}{2015}]{Kastaun_2015}
Kastaun W.,  Galeazzi F.,  2015, \mn@doi [Phys. Rev. D]
  {10.1103/PhysRevD.91.064027}, 91, 064027

\bibitem[\protect\citeauthoryear{Kuroda, Arcones, Takiwaki  \& Kotake}{Kuroda
  et~al.}{2020}]{Kuroda_2020}
Kuroda T.,  Arcones A.,  Takiwaki T.,   Kotake K.,  2020, \mn@doi [The
  Astrophysical Journal] {10.3847/1538-4357/ab9308}, 896, 102

\bibitem[\protect\citeauthoryear{Lattimer \& Swesty}{Lattimer \&
  Swesty}{1991}]{Lattimer:1991nc}
Lattimer J.~M.,  Swesty F.~D.,  1991, \mn@doi [Nucl. Phys. A]
  {10.1016/0375-9474(91)90452-C}, 535, 331

\bibitem[\protect\citeauthoryear{L\"offler et~al.,}{L\"offler
  et~al.}{2012}]{Loffler_2012}
L\"offler F.,  et~al., 2012, \mn@doi [Classical and Quantum Gravity]
  {10.1088/0264-9381/29/11/115001}, 29, 115001

\bibitem[\protect\citeauthoryear{Mart{\'\i}, Ib{\'a}{\~n}ez  \&
  Miralles}{Mart{\'\i} et~al.}{1991}]{Marti:1991wi}
Mart{\'\i} J.~M.,  Ib{\'a}{\~n}ez J.~M.,   Miralles J.~M.,  1991, Phys. Rev. D,
  43, 3794

\bibitem[\protect\citeauthoryear{Modjaz}{Modjaz}{2011}]{Modjaz2011}
Modjaz M.,  2011, \mn@doi [Astronomische Nachrichten]
  {https://doi.org/10.1002/asna.201111562}, 332, 434

\bibitem[\protect\citeauthoryear{Modjaz, Liu, Bianco  \& Graur}{Modjaz
  et~al.}{2016}]{Modjaz_2016}
Modjaz M.,  Liu Y.~Q.,  Bianco F.~B.,   Graur O.,  2016, \mn@doi [The
  Astrophysical Journal] {10.3847/0004-637x/832/2/108}, 832, 108

\bibitem[\protect\citeauthoryear{Moesta et~al.,}{Moesta
  et~al.}{2014}]{Moesta_2014}
Moesta P.,  et~al., 2014, \mn@doi [The Astrophysical Journal]
  {10.1088/2041-8205/785/2/l29}, 785, L29

\bibitem[\protect\citeauthoryear{M{\"o}sta, Ott, Radice, Roberts, Schnetter  \&
  Haas}{M{\"o}sta et~al.}{2015}]{Moesta2015}
M{\"o}sta P.,  Ott C.~D.,  Radice D.,  Roberts L.~F.,  Schnetter E.,   Haas R.,
   2015, \mn@doi [Nature] {10.1038/nature15755}, 528, 376

\bibitem[\protect\citeauthoryear{M\"osta et~al.,}{M\"osta
  et~al.}{2014}]{Moesta2014}
M\"osta P.,  et~al., 2014, \mn@doi [The Astrophysical Journal Letters]
  {10.1088/2041-8205/785/2/L29}, 785, L29

\bibitem[\protect\citeauthoryear{M\"osta, Roberts, Halevi, Ott, Lippuner, Haas
  \& Schnetter}{M\"osta et~al.}{2018}]{Moesta_2018}
M\"osta P.,  Roberts L.~F.,  Halevi G.,  Ott C.~D.,  Lippuner J.,  Haas R.,
  Schnetter E.,  2018, \mn@doi [The Astrophysical Journal]
  {10.3847/1538-4357/aad6ec}, 864, 171

\bibitem[\protect\citeauthoryear{M\"uller, Melson, Heger  \& Janka}{M\"uller
  et~al.}{2017}]{Muller_2017}
M\"uller B.,  Melson T.,  Heger A.,   Janka H.-T.,  2017, \mn@doi [Monthly
  Notices of the Royal Astronomical Society] {10.1093/mnras/stx1962}, 472, 491

\bibitem[\protect\citeauthoryear{Newman \& Hamlin}{Newman \&
  Hamlin}{2014}]{Newman_Hamlin:2014}
Newman W.~I.,  Hamlin N.~D.,  2014, \mn@doi [SIAM Journal on Scientific
  Computing] {10.1137/140956749}, 36, B661

\bibitem[\protect\citeauthoryear{O'Connor \& Ott}{O'Connor \&
  Ott}{2010}]{OConnor_2010}
O'Connor E.,  Ott C.~D.,  2010, \mn@doi [Classical and Quantum Gravity]
  {10.1088/0264-9381/27/11/114103}, 27, 114103

\bibitem[\protect\citeauthoryear{O'Connor \& Ott}{O'Connor \&
  Ott}{2011}]{OConnor_2011}
O'Connor E.,  Ott C.~D.,  2011, \mn@doi [The Astrophysical Journal]
  {10.1088/0004-637X/730/2/70}, 730, 70

\bibitem[\protect\citeauthoryear{Obergaulinger \& Aloy}{Obergaulinger \&
  Aloy}{2021}]{Obergaulinger_2021}
Obergaulinger M.,  Aloy M.~Ã.,  2021, \mn@doi [Monthly Notices of the Royal
  Astronomical Society] {10.1093/mnras/stab295}, 503, 4942

\bibitem[\protect\citeauthoryear{Ott, Burrows, Thompson, Livne  \& Walder}{Ott
  et~al.}{2006}]{Ott_2006}
Ott C.~D.,  Burrows A.,  Thompson T.~A.,  Livne E.,   Walder R.,  2006, \mn@doi
  [The Astrophysical Journal Supplement Series] {10.1086/500832}, 164, 130

\bibitem[\protect\citeauthoryear{Ott et~al.,}{Ott et~al.}{2013}]{Ott_2013}
Ott C.~D.,  et~al., 2013, \mn@doi [The Astrophysical Journal]
  {10.1088/0004-637X/768/2/115}, 768, 115

\bibitem[\protect\citeauthoryear{{Paxton}, {Bildsten}, {Dotter}, {Herwig},
  {Lesaffre}  \& {Timmes}}{{Paxton} et~al.}{2011}]{Paxton2011}
{Paxton} B.,  {Bildsten} L.,  {Dotter} A.,  {Herwig} F.,  {Lesaffre} P.,
  {Timmes} F.,  2011, \mn@doi [\apjs] {10.1088/0067-0049/192/1/3}, \href
  {https://ui.adsabs.harvard.edu/abs/2011ApJS..192....3P} {192, 3}

\bibitem[\protect\citeauthoryear{{Paxton} et~al.,}{{Paxton}
  et~al.}{2013}]{Paxton2013}
{Paxton} B.,  et~al., 2013, \mn@doi [\apjs] {10.1088/0067-0049/208/1/4}, \href
  {https://ui.adsabs.harvard.edu/abs/2013ApJS..208....4P} {208, 4}

\bibitem[\protect\citeauthoryear{Powell, M\"uller, Aguilera-Dena  \&
  Langer}{Powell et~al.}{2023}]{Powell_2023}
Powell J.,  M\"uller B.,  Aguilera-Dena D.~R.,   Langer N.,  2023, \mn@doi
  [Monthly Notices of the Royal Astronomical Society] {10.1093/mnras/stad1292},
  522, 6070

\bibitem[\protect\citeauthoryear{Radice, Galeazzi, Lippuner, Roberts, Ott  \&
  Rezzolla}{Radice et~al.}{2016}]{Radice_2016}
Radice D.,  Galeazzi F.,  Lippuner J.,  Roberts L.~F.,  Ott C.~D.,   Rezzolla
  L.,  2016, \mn@doi [Monthly Notices of the Royal Astronomical Society]
  {10.1093/mnras/stw1227}, 460, 3255

\bibitem[\protect\citeauthoryear{Radice, Perego, Hotokezaka, Fromm, Bernuzzi
  \& Roberts}{Radice et~al.}{2018}]{Radice_2018}
Radice D.,  Perego A.,  Hotokezaka K.,  Fromm S.~A.,  Bernuzzi S.,   Roberts
  L.~F.,  2018, \mn@doi [The Astrophysical Journal] {10.3847/1538-4357/aaf054},
  869, 130

\bibitem[\protect\citeauthoryear{Radice, Bernuzzi, Perego  \& Haas}{Radice
  et~al.}{2022}]{Radice:2022}
Radice D.,  Bernuzzi S.,  Perego A.,   Haas R.,  2022, \mn@doi [Monthly Notices
  of the Royal Astronomical Society] {10.1093/mnras/stac589}, 512, 1499

\bibitem[\protect\citeauthoryear{Ressler, Tchekhovskoy, Quataert  \&
  Gammie}{Ressler et~al.}{2017}]{Ressler_2017}
Ressler S.~M.,  Tchekhovskoy A.,  Quataert E.,   Gammie C.~F.,  2017, \mn@doi
  [Monthly Notices of the Royal Astronomical Society] {10.1093/mnras/stx364},
  467, 3604

\bibitem[\protect\citeauthoryear{Ruiz, Hilditch  \& Bernuzzi}{Ruiz
  et~al.}{2011}]{Milton:2011}
Ruiz M.,  Hilditch D.,   Bernuzzi S.,  2011, \mn@doi [Phys. Rev. D]
  {10.1103/PhysRevD.83.024025}, 83, 024025

\bibitem[\protect\citeauthoryear{Schnetter, Brandt, Cupp, Haas, M\"osta  \&
  Shankar}{Schnetter et~al.}{2022}]{schnetter_2022_zenodo}
Schnetter E.,  Brandt S.,  Cupp S.,  Haas R.,  M\"osta P.,   Shankar S.,  2022,
  CarpetX, \mn@doi{10.5281/zenodo.6131529}, \url
  {https://doi.org/10.5281/zenodo.6131529}

\bibitem[\protect\citeauthoryear{Shankar, M\"osta, Brandt, Haas, Schnetter  \&
  de Graaf}{Shankar et~al.}{2023}]{Shankar2023}
Shankar S.,  M\"osta P.,  Brandt S.~R.,  Haas R.,  Schnetter E.,   de Graaf Y.,
  2023, \mn@doi [Classical and Quantum Gravity] {10.1088/1361-6382/acf2d9}, 40,
  205009

\bibitem[\protect\citeauthoryear{Shibagaki, Kuroda, Kotake, Takiwaki  \&
  Fischer}{Shibagaki et~al.}{2024}]{Shibagaki_2024}
Shibagaki S.,  Kuroda T.,  Kotake K.,  Takiwaki T.,   Fischer T.,  2024,
  \mn@doi [Monthly Notices of the Royal Astronomical Society]
  {10.1093/mnras/stae1361}, 531, 3732

\bibitem[\protect\citeauthoryear{Shibata, Liu, Shapiro  \& Stephens}{Shibata
  et~al.}{2006}]{Shibata_2006}
Shibata M.,  Liu Y.~T.,  Shapiro S.~L.,   Stephens B.~C.,  2006, \mn@doi [Phys.
  Rev. D] {10.1103/PhysRevD.74.104026}, 74, 104026

\bibitem[\protect\citeauthoryear{Shu}{Shu}{1998}]{shu:98}
Shu C.-W.,  1998, Lecture Notes in Mathematics, 1697, 325

\bibitem[\protect\citeauthoryear{Takiwaki \& Kotake}{Takiwaki \&
  Kotake}{2011}]{Takiwaki_2011}
Takiwaki T.,  Kotake K.,  2011, \mn@doi [The Astrophysical Journal]
  {10.1088/0004-637X/743/1/30}, 743, 30

\bibitem[\protect\citeauthoryear{Toro}{Toro}{1999}]{toro:99}
Toro E.~F.,  1999, {R}iemann {S}olvers and {N}umerical {M}ethods for {F}luid
  {D}ynamics.
Springer, Berlin

\bibitem[\protect\citeauthoryear{Winteler, K\"appeli, Perego, Arcones, Vasset,
  Nishimura, Liebend\"orfer  \& Thielemann}{Winteler
  et~al.}{2012}]{Winteler_2012}
Winteler C.,  K\"appeli R.,  Perego A.,  Arcones A.,  Vasset N.,  Nishimura N.,
  Liebend\"orfer M.,   Thielemann F.-K.,  2012, \mn@doi [The Astrophysical
  Journal Letters] {10.1088/2041-8205/750/1/L22}, 750, L22

\bibitem[\protect\citeauthoryear{Woosley \& Bloom}{Woosley \&
  Bloom}{2006}]{Woosley_2006}
Woosley S.,  Bloom J.,  2006, \mn@doi [Annual Review of Astronomy and
  Astrophysics] {https://doi.org/10.1146/annurev.astro.43.072103.150558}, 44,
  507

\makeatother
\end{thebibliography}

\bsp	
\label{lastpage}
\end{document}